# Anomalous high-temperature superconductivity in $YH_6$

Ivan A. Troyan,[1,&] Dmitrii V. Semenok,[2,&,*] Alexander G. Kvashnin,[2,&,*] Andrey V. Sadakov,[3] Oleg A. Sobolevskiy,[3] Vladimir M. Pudalov,[3] Anna G. Ivanova,[1] Vitali B. Prakapenka,[4] Eran Greenberg,[4] Alexander G. Gavriliuk,[1,5] Viktor V. Struzhkin,[6] Aitor Bergara,[10,11,12] Ion Errea,[11,12,13] Raffaello Bianco,[11] Matteo Calandra,[14,15,16] Francesco Mauri,[16,17] Lorenzo Monacelli,[16,17] Ryosuke Akashi,[18] and Artem R. Oganov[2,8,9,*]

[1] Shubnikov Institute of Crystallography, Federal Scientific Research Center Crystallography and Photonics, Russian Academy of Sciences, Moscow 119333, 59 Leninskii Prospect, Russia
[2] Skolkovo Institute of Science and Technology, Skolkovo Innovation Center, 3 Nobel Street, Moscow 121205, Russia
[3] P. N. Lebedev Physical Institute, Russian Academy of Sciences, Moscow 119991, Russia
[4] Center for Advanced Radiation Sources, The University of Chicago, 5640 South Ellis Avenue, Chicago, Illinois 60637, USA
[5] IC RAS Institute for Nuclear Research, Russian Academy of Sciences, Moscow, 117312 Russia
[6] Center for High Pressure Science and Technology Advanced Research, Shanghai 201203, China
[8] Moscow Institute of Physics and Technology, 9 Institutsky Lane, Dolgoprudny 141700, Russia
[9] International Center for Materials Discovery, Northwestern Polytechnical University, Xi'an 710072, China
[10] Departamento de Física de la Materia Condensada, University of the Basque Country (UPV/EHU), 48080 Bilbao, Basque Country, Spain
[11] Centro de Física de Materiales CFM, CSIC-UPV/EHU, Paseo Manuel de Lardizabal 5, 20018 Donostia, Basque Country, Spain
[12] Donostia International Physics Center (DIPC), Manuel Lardizabal pasealekua 4, 20018 Donostia, Basque Country, Spain
[13] Fisika Aplikatua 1 Saila, University of the Basque Country (UPV/EHU), Europa Plaza 1, 20018 Donostia, Basque Country, Spain
[14] Departimento di Fisica, Universit di Trento, Via Sommarive 14, 38123 Povo, Italy
[15] Sorbonne Université, CNRS, Institut des Nanosciences de Paris, UMR7588, F-75252 Paris, France
[16] Graphene Labs, Fondazione Istituto Italiano di Tecnologia, Via Morego, I-16163 Genova, Italy
[17] Dipartimento di Fisica, Università di Roma Sapienza, Piazzale Aldo Moro 5, I-00185 Roma, Italy
[18] University of Tokyo, 7-3-1 Hongo, Bunkyo, Tokyo 113-8654, Japan

## Abstract

Pressure-stabilized hydrides are a new rapidly growing class of high-temperature superconductors which is believed to be described within the conventional phonon-mediated mechanism of coupling. Here we report the synthesis of yttrium hexahydride $Im\bar{3}m$-$YH_6$ that demonstrates the superconducting transition with $T_C$ ~ 224 K at 166 GPa, much lower than the theoretically predicted (>270 K). The measured upper critical magnetic field $B_{c2}(0)$ of $YH_6$ was found to be 116–158 T, which is 2–2.5 times larger than the calculated value. A pronounced shift of $T_C$ in yttrium deuteride $YD_6$ with the isotope coefficient 0.4 supports the phonon-assisted superconductivity. Current-voltage measurements showed that the critical current $I_C$ and its density $J_C$ may exceed 1.75 A and 3500 A/mm$^2$ at 0 K, respectively, which is comparable with the parameters of commercial superconductors, such as NbTi and YBCO. The superconducting density functional theory (SCDFT) and anharmonic calculations suggest unusually large impact of the Coulomb repulsion in this compound. The results indicate notable departures of the superconducting properties of the discovered $YH_6$ from the conventional Migdal-Eliashberg and Bardeen–Cooper–Schrieffer theories.

**Keywords:** yttrium hydrides, superconductivity, USPEX, high pressure, SCDFT, SSCHA

**Highlights**

- High-$T_C$ superconductivity in $Im\bar{3}m$-$YH_6$ at 224 K and 166 Gigapascals
- Upper critical magnetic field of $YH_6$ is 116–158 Tesla at 0 K
- Critical current density (~3500 A/mm$^2$) in yttrium hexahydride may exceed parameters of YBCO and NbTi.
- Notable discrepancies between the $T_C$ and $B_{C2}(0)$ predicted within BCS model, and the experimental values

# Introduction

Room-temperature superconductivity has been an unattainable dream and subject of speculative discussions for a long time, but times change. The theoretical prediction of record high-temperature superconductor $LaH_{10}$[1] followed by the experimental confirmation of its critical temperature $T_C$ ~ 250–260 K[2–4] has opened a new field in high-pressure physics devoted to the investigation of superconducting metal hydrides. Recent successful synthesis of previously predicted superconducting $BaH_{12}$,[5] $ThH_{10}$,[6] $UH_7$ and $UH_8$,[7] $CeH_9$,[8] $PrH_9$,[9] and $NdH_9$[10] motivated us to perform an experimental study of the Y–H system to find previously predicted potential room-temperature superconductor $Im\bar{3}m$-$YH_6$, stable in pressure range of 110–300 GPa.[1,11,12]

The outstanding superconducting properties combined with a relatively low predicted stabilization pressure of about 110 GPa[11] make yttrium hexahydride very interesting. Starting from 2015, the stability, conditions and physical properties of $YH_6$, having a sodalite-like crystal structure similar to another predicted hexahydride $Im\bar{3}m$-$CaH_6$, have been studied in several works.[1,11,12] In 2015, Li et al.[11] have predicted the stability of $Im\bar{3}m$-$YH_6$ at pressures over 110 GPa. Solving the Migdal–Eliashberg (ME) equations numerically, they found a superconducting transition temperature $T_C$ = 251–264 K at 120 GPa ($\mu^*$ = 0.1−0.13), with the electron-phonon coupling (EPC) coefficient $\lambda$ reaching 2.93. In the study of the physical properties and superconductivity of $Im\bar{3}m$-$YH_6$ by Heil et al. in 2019,[13] the most detailed so far, the calculations were made using the fully anisotropic Migdal–Eliashberg theory (as implemented in the EPW code) with Coulomb corrections. They have found that an almost isotropic superconducting gap in $YH_6$ is caused by a uniform distribution of the coupling over the states of both Y and H sublattices and have predicted the critical temperature $T_C$ = 290 K at 300 GPa[13]. A summary of the previous results of theoretical studies of $Im\bar{3}m$-$YH_6$ is shown in Supplementary Table S6 and Fig. S11.

Following the theoretical predictions, in this work we report experimental study of the superconducting properties of yttrium hexahydride $Im\bar{3}m$-$YH_6$, synthesized together with $YH_7$ and $YH_4$, after laser heating yttrium samples compressed to 166–172 GPa in the ammonia borane ($NH_3BH_3$) medium in the diamond anvil cells.

# Results and discussion

In the first part of this work we focused on the experimental verification of stability and superconductivity of $Im\bar{3}m$-$YH_6$ and on the calculation of some physical properties that have not been analyzed before.

The high-pressure synthesis was carried out with ammonia borane as a source of hydrogen, following the technique that has shown good results in previous studies.[2,3,6,9,10] We prepared three diamond anvil cells (DACs) with 50 μm culets (K1, M1, and M3), where pure yttrium metal was loaded into sublimated ammonia borane and compressed to 166–172 GPa. The pulsed laser heating technique was used to heat samples at 2400 K ($10^5$ pulses, 1 μs pulse width, 10 kHz) that resulted in the formation of three compounds $Im\bar{3}m$-$YH_6$ and $YH_4$ or $YH_{7+x}$ in all DACs.

The results of the synthesis are strongly dependent on pressure and temperature conditions. In DACs K1 and M1 (the X-ray diffraction (XRD) patterns are shown in Figure 1 and Supplementary Fig. S7 and S8), the laser heating of the samples at 166 GPa yielded a complex mixture of products with predominant $Im\bar{3}m$-$YH_6$ and, probably, $P1$-$YH_7$ or pseudocubic $Imm2$-$YH_7$ (Supplementary Fig. S7), which have been found using the USPEX structure search.[14–16]

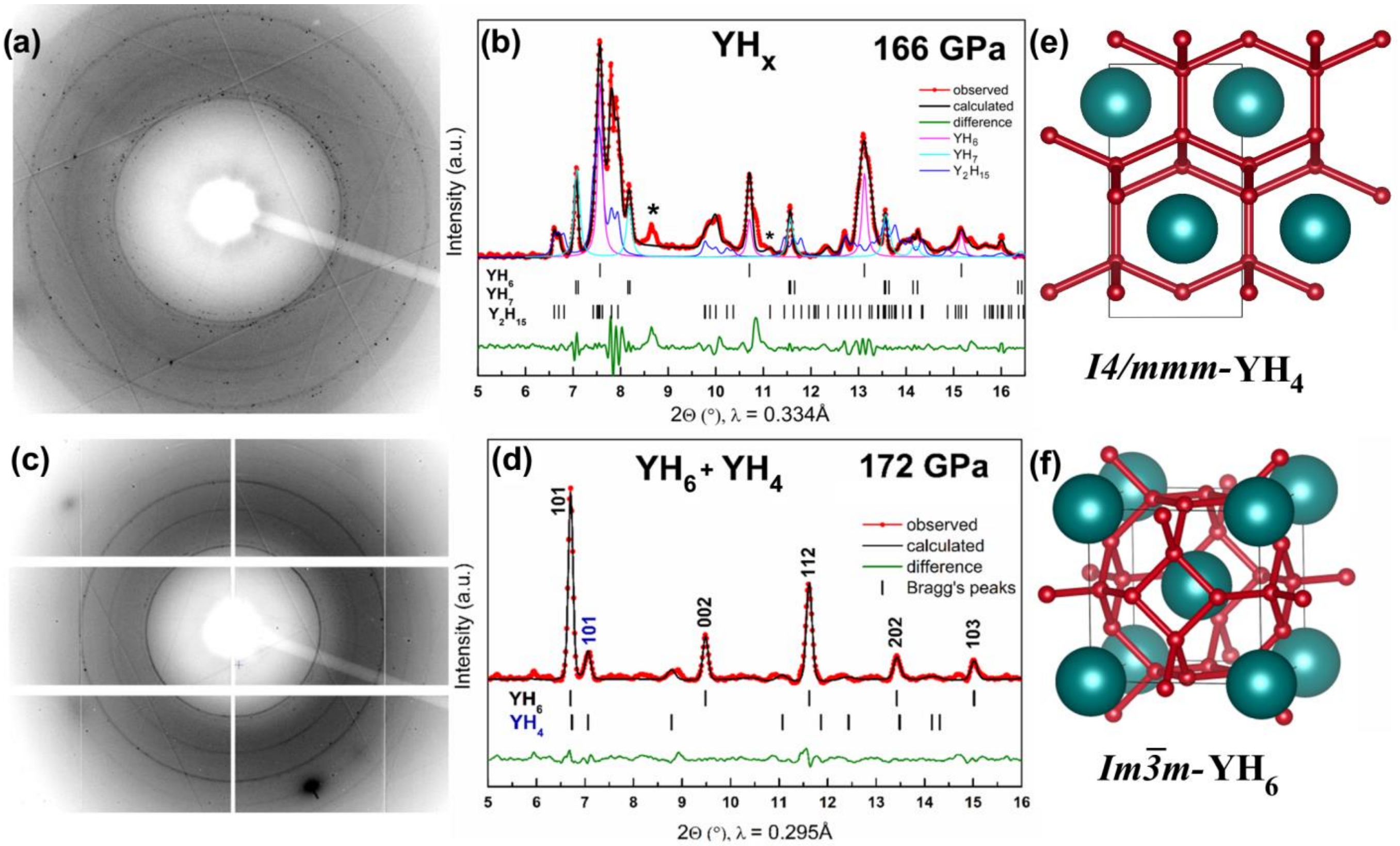

**Figure 1.** (a) The X-ray diffraction (XRD) pattern of the sample in DAC K1 at 166 GPa collected at a wavelength of $\lambda = 0.3344$ Å. (b) The Le Bail refinements of $Im\bar{3}m$-$YH_6$, $Imm2$-$YH_7$, and $P1$-$YH_{7+x}$ ($x = \pm 0.5$) at 166 GPa. Unidentified reflections are marked by asterisks. (c) XRD pattern of the M3 sample at 172 GPa collected with $\lambda = 0.2952$ Å. (d) The Le Bail refinements of $Im\bar{3}m$-$YH_6$ and $I4/mmm$-$YH_4$. The experimental data, fitted line, and residues are shown in red, black, and green, respectively. (e-f) Crystal structures of $YH_4$ and $YH_6$.

At higher pressures (172–180 GPa, DAC M3), a much simpler XRD pattern was observed (Figure 1c, d), with peaks only from the $Im\bar{3}m$-$YH_6$ and distorted $I4/mmm$-$YH_4$ phases. The experimental lattice parameters and volumes of synthesized $Im\bar{3}m$-$YH_6$ are given in the Table 1 (for $YH_4$ and $YH_7$, see Supplementary Tables S3 and S5). All Y–H phases were also theoretically examined for the dynamic and mechanical stability according to the Born criteria ($C_{11} - C_{12} > 0$, $C_{11} + 2C_{12} > 0$, $C_{44} > 0$), the obtained results are presented in Supplementary Tables S10–S12.

**Table 1.** The experimental ($a$, $V$) and predicted ($a_{DFT}$, $V_{DFT}$) lattice parameters and volumes of $Im\bar{3}m$-$YH_6$ ($Z = 2$).

| DAC | Pressure, GPa | $a$, Å | $V$, Å$^3$ | $a_{DFT}$, Å | $V_{DFT}$, Å$^3$ |
|---|---|---|---|---|---|
| M1 | 166 | 3.578(3) | 45.82 | 3.573 | 45.62 |
| K1 | 168 | 3.582(3) | 45.91 | 3.565 | 45.31 |
| M3 | 172 | 3.571(2) | 45.53 | 3.557 | 45.02 |
| M3 | 177 | 3.565(9) | 45.34 | 3.551 | 44.79 |
| M3 | 180 | 3.559(8) | 45.07 | 3.546 | 44.58 |

To estimate the thermodynamic stability and the possibility of formation of these hydrides at the experimental pressure-temperature conditions, we carried out searches for stable Y–H compounds using the evolutionary algorithm USPEX[14,15,17] at 150, 200, 250, and 300 GPa. The results of the computational predictions at 0, 500, 1000, and 2000 K (with the zero-point energy (ZPE) included at the harmonic level) and a pressure of 150 GPa are shown in Figure 2 (for other pressures, see Supplementary Fig. S2–S6).

At 150 GPa and 0 K, with the ZPE contribution taken into account, the only stable hydrides are $Fm\bar{3}m$-YH and $YH_3$, *I4/mmm*-$YH_4$, *P*1-$YH_7$ (*Imm*2-$YH_7$ is a bit less stable), and pseudohexagonal $P\bar{1}$-$YH_9$, whereas $Im\bar{3}m$-$YH_6$ is metastable, lying 30 meV/atom above the convex hull (Figure 2a). The distorted hydrogen sublattice in $P\bar{1}$-$YH_9$ leads to a lower enthalpy of formation compared with previously proposed $P6_3/mmc$-$YH_9$,[18] which has an ideal hexagonal structure (Supplementary Fig. S1). The cubic modification of $YH_9$ with space group $F\bar{4}3m$ (isostructural with $PrH_9$)[9] is more stable than $P6_3/mmc$-$YH_9$[18] (Figure 2a).

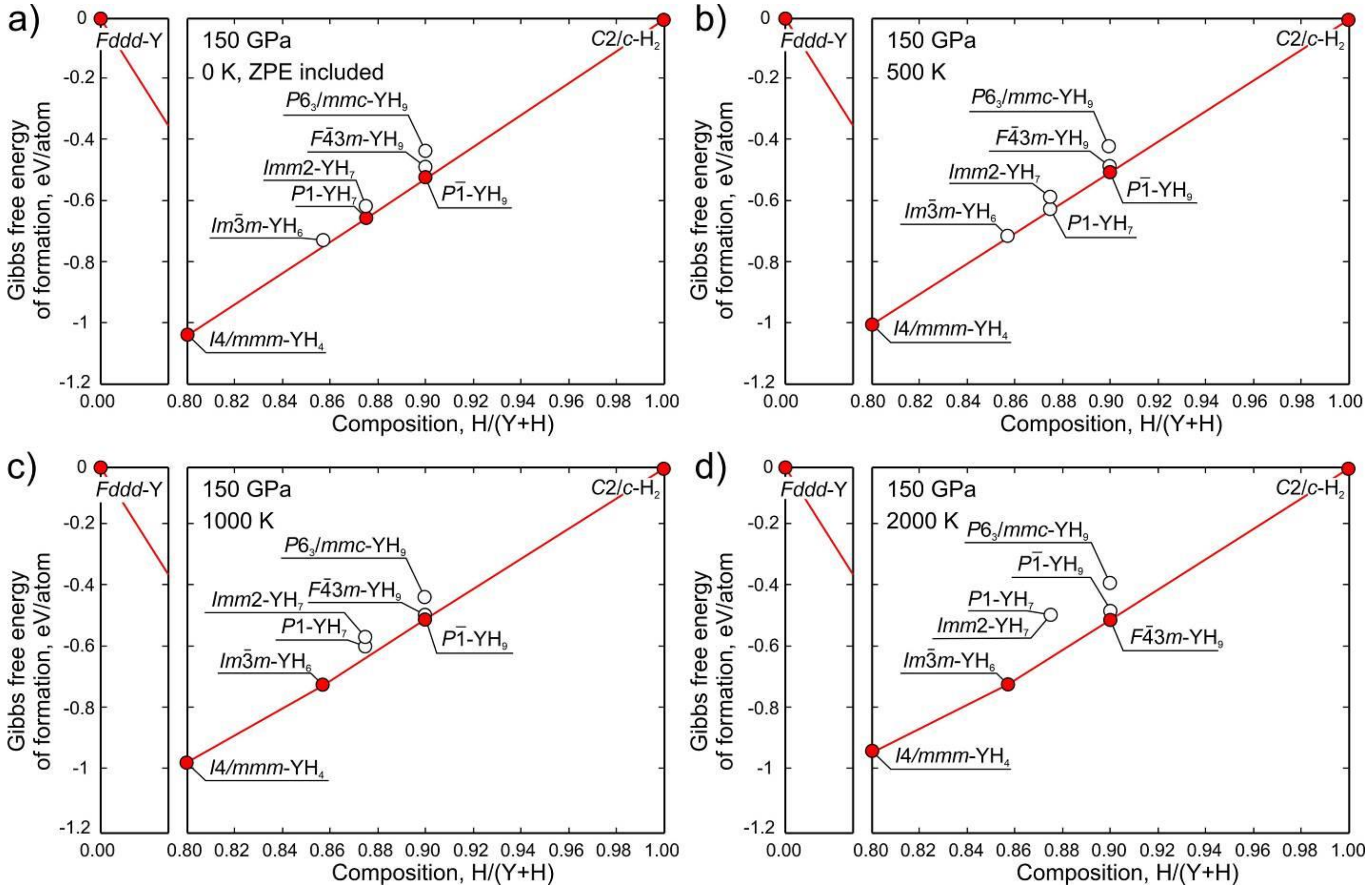

**Figure 2.** Calculated convex hulls of the Y–H system at 150 GPa and (a) 0 K, (b) 500 K, (c) 1000 K, and (d) 2000 K.

As the temperature rises, the separation of the $YH_7$ from the convex hull also increases to 110 meV/atom at 2000 K (Figure 2d). Laser heating of the samples above 1000 K (Figure 2c) leads to the stabilization of $Im\bar{3}m$-$YH_6$ and transformation of the pseudohexagonal $P\bar{1}$-$YH_9$ to the cubic modification of $YH_9$, which becomes stable at temperatures above 1500 K (Figure 2d). Calculations at 200 and 250 GPa (Supplementary Fig. S5 and S6) show that both *P*1-$YH_7$ and $Im\bar{3}m$-$YH_6$, which below 150 GPa is a metastable "frozen" phase, can form simultaneously, in accordance with the experimental data. In contrast to early predictions[1,11,12], our computations, considering *P*1-$YH_7$ and $P\bar{1}$-$YH_9$, show that yttrium hexahydride stabilizes at 100-150 GPa only due to the entropy factor.

It is interesting that $Fm\bar{3}m$-$YH_{10}$ is thermodynamically metastable at 200–250 GPa and 0–2000 K, with the Gibbs free energy of formation at least 18 meV/atom above the convex hull because of the existence of $YH_9$ (Supplementary Figure 25 and S6). This may explain the failure to synthesize $Fm\bar{3}m$-$YH_{10}$ at 243 GPa.[18] On the other hand, the stabilization of $F\bar{4}3m$-$YH_9$ may shed light on the recent detection of superconductivity in $YH_x$ at 262 K by the group of Ranga Dias.[19]

To measure the superconducting transition temperature of synthesized yttrium hexahydride, all DACs were equipped with four Ta/Au electrodes. We used the DACs with a 50 μm culet bevelled to 300 μm at 8.5°. Four Ta electrodes (~200 nm thick) with a gold plating (~80 nm) were sputtered on the diamond anvil.

The composite gaskets consisting of a tungsten ring and a $CaF_2$/epoxy mixture were used to isolate the electrical leads. To measure the isotope effect in $YD_6$, a similar cell loaded with $ND_3BD_3$ was prepared.

An yttrium sample with a thickness of ~1–2 μm was sandwiched between the electrodes and ammonia borane in the gasket hole with a diameter of 20 μm. In the DAC M3, the electrodes were in short-circuit with the tungsten gasket, therefore no resistivity measurements were made. The temperature dependence of the resistance is shown in Figure 3. In each cooling cycle, the pressure increased because of the thermal expansion of the DAC materials.

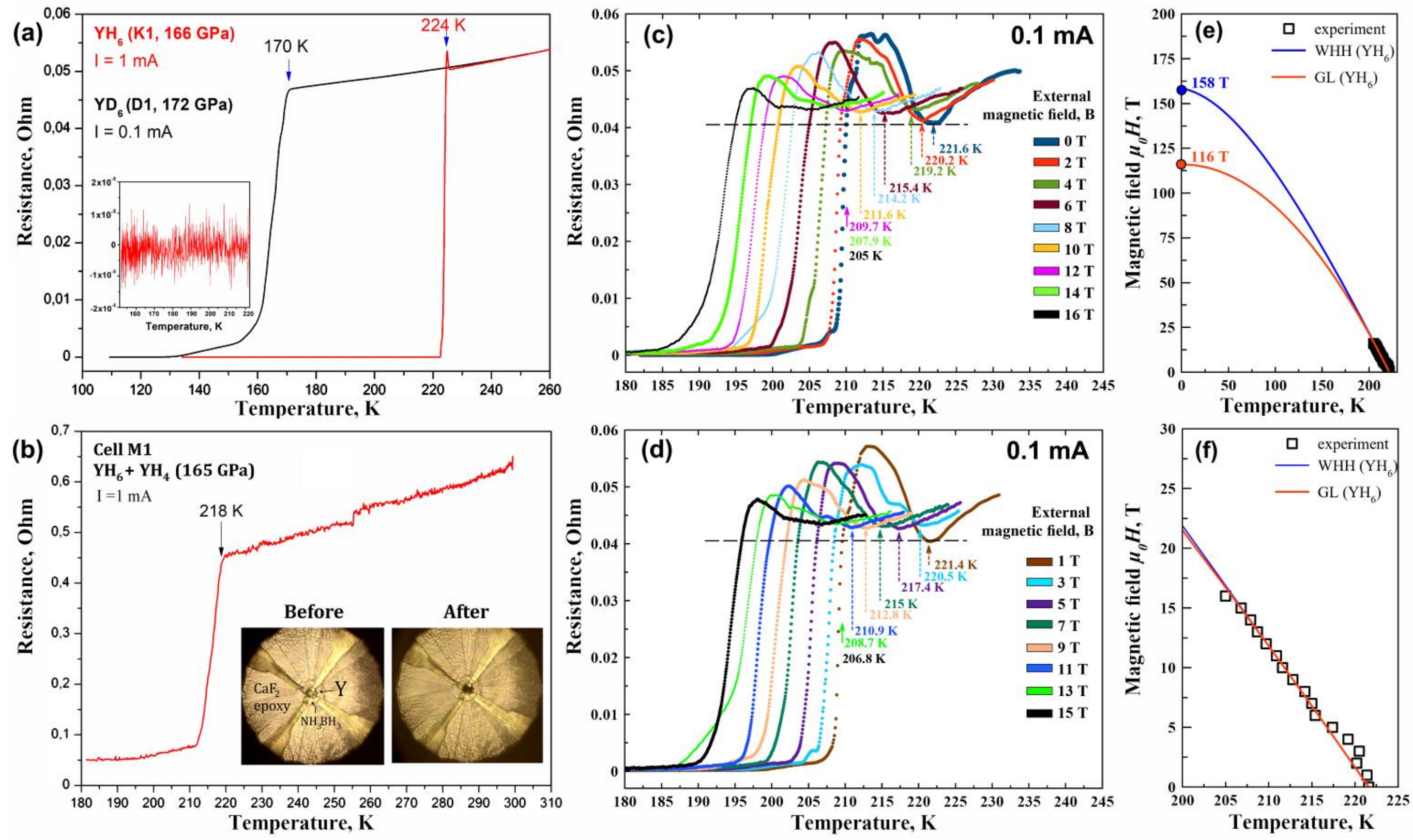

**Figure 3.** Superconducting transitions in the $Im\overline{3}m$-$YH_6$: (a) the temperature dependence of the electrical resistance $R(T)$ in the $YH_6$ (DAC K1) and $YD_6$ (DAC D1). Inset: the resistance drops to zero after cooling below $T_C$; (b) temperature dependence of the electrical resistance in DAC M1. A 9-fold decrease is observed; (c, d) dependence of the electrical resistance on the external magnetic field (0–16 T) at 183 GPa and a current of 0.1 mA for (c) even and (d) odd values of the magnetic field. Due to presence several hydride phases in the sample the superconducting transition in $YH_6$ can be observed as an upward feature of the R(T,H) curves due to the shunting effect in the fine-grained samples. The critical temperatures were determined at the onset of the resistance jump; (e) the upper critical magnetic field estimated using the Werthamer–Helfand–Hohenberg theory[20] and the Ginzburg–Landau[21] theory; (f) the dependence of the critical temperature $T_C$ ($YH_6$) on the applied magnetic field.

Two slightly different superconducting transitions in the $YH_6$ with $T_C$ of 224 K (Figure 3a) and 218 K (Figure 3b) were observed in the DACs K1 and M1, respectively. In the K1 cell, the electrical resistance dropped sharply to zero (from 50 mΩ to 5 μΩ, with $\Delta T_C$ ~ 1–2 K) due to good location of the sample, whereas in the M1 cell the electrical resistance did not disappear completely because of the presence of additional phases (Supplementary Fig. S7). The $YD_6$ sample synthesized using deuterated ammonium borane demonstrates superconducting transition at 170 K, which corresponds to the isotope coefficient $\alpha_{exp} = 0.4$, lower than the BCS theory gives (~0.5).

The analysis of the electronic and superconducting properties of the tetragonal $YH_4$ and pseudocubic *Imm*2-$YH_7$ shows that the *I*4/*mmm*-$YH_4$ is a metal with a significantly lower calculated critical temperature (≤94 K) compared with $YH_6$. Another possible product of the synthesis in DAC M1, *Imm*2-$YH_7$, has a pronounced pseudogap in the electronic density of states $N(E)$, leading to a quite low density of states at the Fermi level $N(E_F)$, and, as a result, low predicted $T_C$ of 32–43 K (Supplementary Table S7, Fig. S14). Other possible crystal modifications of $YH_7$, such as *Cc* or *P*1, have even lower $N(E_F)$ and $T_C$. Thus, the presence of these phases has practically no effect on the superconducting transition in the yttrium hexahydride.

In $YH_6$, the dependence of the $T_C$ ($YH_6$) on the magnetic induction $B = \mu_0 H$ was measured at 183 GPa in the range of 0–16 T (Figure 3c, d) and interpolated using the Ginzburg–Landau[21] and the Werthamer–Helfand–Hohenberg (WHH)[20] models simplified by Baumgartner.[22] Around 200 K, an almost linear dependence of the $T_C(B)$ with a gradient $dB_{c2}/dT \approx -1$ T/K was observed (Figure 3f). The experimentally found upper critical field $\mu_0 H_{c2}(0)$ for $YH_6$ is 116–158 T, in agreement with the results of Kong et al.[18] These data allows us to estimate $N(E_F)(1+\lambda)$ factor in the interpolation formula proposed by Carbotte[23] for the $B_{c2}(0)$ of conventional superconductors (Supplementary Table S7) at 7.2–13.3 $eV^{-1}f.u.^{-1}$.

One of distinguishing features of superconductors is the existence of an upper limit of the current density ($J_C$) at which superconductivity disappears. The critical currents and the voltage–current (V–I) characteristics for the $YH_6$ sample were investigated in the range of $10^{-4}$–$10^{-2}$ A in external magnetic fields after further compression to 196 GPa (Figure 4a, b). The critical current density was estimated on the basis of the facts that the size of the sample cannot exceed the size of the culet (50 μm), and the thickness of the sample is smaller than the thickness of the gasket before the cell is loaded, ~10 μm. A zero-field cooling (ZFC) shows that the critical current density in $YH_6$ exceeds $2\times10^7$ A/m$^2$ at $T = 190$ K. To compare critical currents in magnetic fields at low temperatures we used a single vortex model. Critical current density in magnetic fields may be defined as the current that creates strong enough force to de-pin a vortex or a bundle of vortices. There are two possible sources of pinning: non-superconducting (normal) particles embedded in the superconducting matrix leading to a scattering of electrons, so called "dl-pinning", or pinning provided by spatial variations of the Ginzburg parameter ($\kappa = \lambda/\xi$) associated with fluctuations in the transition temperature $T_C$, so called "d$T_C$-pinning" (or dk-pinning).

Analysis of the pinning force ($F_p = B\cdot I_C$) dependence on magnetic field (Supplementary Fig. S13a) shows that according to Dew-Hughes [24] the dominant type of pinning in $YH_6$ is "dl-pinning". This allows us to extrapolate $I_C$(T) data to low temperatures within the single vortex model $J_C = J_{c0}(1 - T/T_C)^{5/2}(1 + T/T_C)^{-1/2}$ which is suitable for rather low fields of several Tesla in the whole temperature range (see equation (8) in Ref. [25]). The extrapolation shows that at 4.2 K the critical current $I_C$ in the sample can reach 1.75 A and the critical current density $J_C$ may exceed 3500 A/mm$^2$ (Figure 4c). However, using the Ginzburg–Landau model,[21] $J_C = J_{c0}(1 - T/T_C)^{3/2}$, gives lower values: the maximum critical current $I_C$ in the sample ~1 A, and the maximum critical current density $J_C$ is about 2000 A/mm$^2$ (Supplementary Fig. S13b). These values of $J_C$ are comparable with the parameters of commercial superconducting materials like NbTi and YBCO [26] (Figure 4d), which opens prospects of using superhydrides in electronic devices.

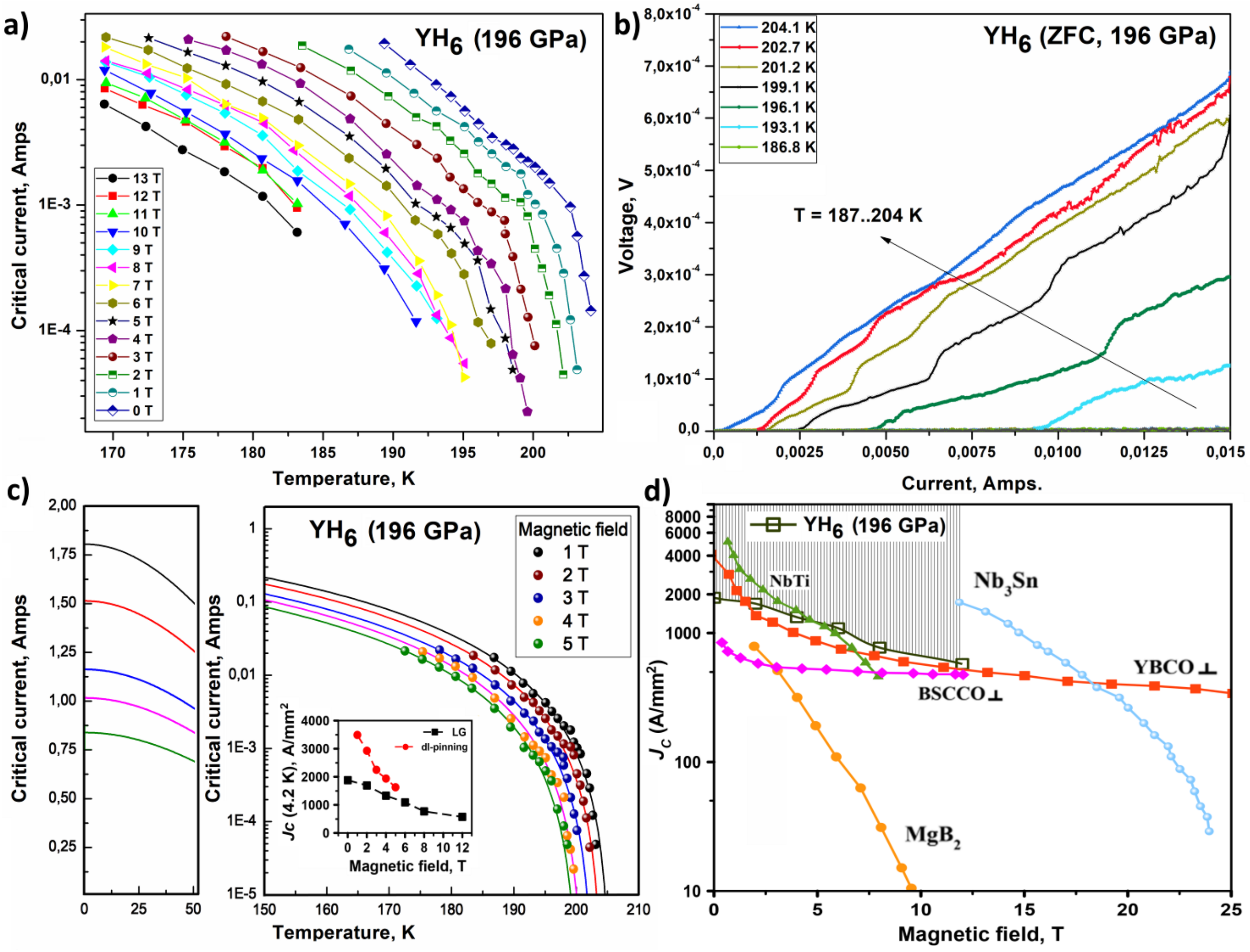

**Figure 4**. Dependence of the critical current on the temperature and external magnetic fields (0–13 T) in $Im\bar{3}m$-$YH_6$ at 196 GPa. (a) The critical current at different magnetic fields near $T_C$ (defined below 50% resistance drop). (b) The voltage–current characteristics of the $YH_6$ sample near $T_C$. (c) Extrapolation of the temperature dependence of the critical current using the single vortex model [24] $J_C = J_{c0}(1 - T/T_C)^{5/2}(1 + T/T_C)^{-1/2}$; inset: dependencies of the critical current density at 4.2 K on the magnetic field. (d) The critical current densities $J_C$ of various industrial superconducting wires and the $YH_6$ (shaded area) at 4.2 K. The lower bound of the critical current density of $YH_6$ was calculated assuming the maximum possible cross section of the sample of 10×50 μm$^2$.

It is interesting to compare the experimentally obtained $T_C$ with theoretical calculations based on the Bardeen–Cooper–Schrieffer[25–27] and the Migdal–Eliashberg[30,31] theories. In early theoretical works, estimated $T_C$ varies from 250 to 285 K,[1,13] which is quite far from the experimental values (224–226 K). Considering that the studied pressure range of 165–180 GPa has not been covered in previous papers, we carried out a series of calculations of the superconducting properties of $Im\bar{3}m$-$YH_6$ at a fixed pressure of 165 GPa (Supplementary Tables S7–S9) within a semiphenomenological Migdal–Eliashberg[30,31] approach, where the electron-electron Coulomb repulsion is accounted by one empirical parameter, the Coulomb pseudopotential μ*, and a parameter-free superconducting density functional theory (SCDFT)[32,33].

Numerical solution of the isotropic Migdal–Eliashberg equations[31] within the standard range of μ*(cut-off frequency is 6 Ry) = 0.15–0.1 yields a $T_C$ = 261–272 K, which is substantially higher than the experimental value for $YH_6$. This significant theoretical overestimation of the critical temperature calculated within the harmonic approach motivated us to include the effects of anharmonicity by performing calculations of the phonon band structure and Eliashberg function $\alpha^2F(\omega)$ using the stochastic self-consistent harmonic approximation (SSCHA)[34–36] (Supplementary Fig. S21 and S15), which is a non-perturbative variational method to consider anharmonic effects.

At a pressure of 165 GPa, the calculations with the anharmonic $\alpha^2F(\omega)$ show a decrease in the EPC coefficient of $YH_6$ from 2.24 to 1.71 and an increase of $\omega_{log}$ from 929 to 1333 K due to hardening of the optical phonon modes. The overall influence of anharmonicity on the critical temperature is a decrease by 25 K, and resulting $T_C$ is 236–247 K, which is still higher than the experimental values. To match the experimental data, including isotope coefficient $\alpha_{exp}$ = 0.4, unusually high values of the Coulomb pseudopotential $\mu$*(6 Ry) = 0.19–0.22 are necessary. Estimations for $YH_6$ on the basis of DFT-calculated $N(E_F)$, $\lambda$, and $\omega_{log}$ give an expected upper critical magnetic field $\mu_0 H_C(0)$ ~ 60 T, a superconducting gap of 48 meV, a coherence length $\xi_{BCS} = 0.5\sqrt{h/\pi e H_{c2}}$ = 23 Å, and $N(E_F)(1+\lambda) = 1.92$ (Supplementary Table S7).

The experimentally found upper critical magnetic field $\mu_0 H_{c2}(0)$ exceeds 110 T (and can reach 158 T in the WHH model), which is more than 2–2.5 times higher than the value predicted within the BCS theory. In other words, the term $N(E_F)(1+\lambda)$, related to the Sommerfeld constant, is at least 4 times higher than follows from DFT calculations, and $\xi_{exp}$ is 14-17 Å. In this regard, it is curious that possible deviation from the BCS theory of superconductivity in $YH_6$ was recently noted in Ref. [37] on the basis of the $T_C$-$T_F$ Uemura plot [38] for superhydrides. A similar disagreement, but less pronounced, is also observed for $Fm\bar{3}m$-$LaH_{10}$,[2] where the experimental $\mu_0 H_{c2}(0)$ exceeds the predicted one by ~30–40%.

The SCDFT calculations, which incorporate the phonon-mediated pairing, mass renormalization and pair-breaking Coulomb repulsion under the retardation effect without empirical parameters, such as $\mu^*$ in the Migdal-Eliashberg equations, reveal another anomaly of $YH_6$. Solving the SCDFT gap equation with the anharmonic $\alpha^2F(\omega)$ (Supplementary Information equation (S1)) at 165 GPa yields $T_C$ = 160 K and Coulomb potential $\mu$ = 0.187, with an error bar of ~2.5% originating from the random sampling step in solving the equation.[33,39] This value is about 100 K lower than the one found using the Migdal–Eliashberg calculations and 64 K (~28%) below the experimental value (224 K). A similar discrepancy has been reported in $LaH_{10}$[40] and within ab initio Eliashberg theory [41], but in those cases the difference was ~20-30 K. Apart from this, the 64 K underestimation still indicates anomalously large impact of the Coulomb repulsion and implies something beyond the conventional phonon-mediated superconductivity that boosts the critical temperature up to the experimentally observed value.

Within the conventional mechanism, we can try to explain the discrepancy using at least two hypotheses. First, it could be achieved in the fully anisotropic Eliashberg equation.[42,43] In the work of Heil et al.,[13] at low temperatures the gap function showed a dispersion of width of approximately 30 meV, indicating that the pairing strength depends significantly on the band index and wavenumber. As seen in multiband superconductor $MgB_2$, averaging approaches using $\alpha^2F(\omega)$ generally yields smaller $T_C$'s for systems with such band and wavenumber dependences.[44] Although the SCDFT gap equation presumably incorporates the effects included in the Eliashberg equations, they both can give a little different $T_C$ values, as has been pointed out for $LaH_{10}$.[43] Also, large phonon energy scales in the hydride could make relevant the higher order electron–phonon coupling effects beyond the Born–Oppenheimer and the Migdal approximations. Sano et al.[45] demonstrated that in superconducting sulfur hydride $H_3S$, the Debye–Waller correction [46] to the electronic band structure (including the finite spread of the ionic sites) and the vertex correction (including the higher order perturbation of the self-energy due to the electron–phonon interaction) both change the calculated value of $T_C$ by several tens of Kelvins. It would be interesting to explore such effects in $YH_6$.

# Conclusions

In this research, the novel high-$T_C$ superconductor $Im\bar{3}m$-$YH_6$ was discovered together with $I4/mmm$-$YH_4$ and $YH_7$ at pressures of 160–196 GPa, confirming theoretical predictions.[1] The low-symmetry molecular yttrium hydride $P1$-$YH_{7+x}$ ($x = \pm 0.5$) was found to cause complex XRD patterns at 166 GPa. The measured critical temperature of $Im\bar{3}m$-$YH_6$ is 224 K, which is unexpectedly lower than the theoretically predicted value (>273 K).[9] The observed upper critical magnetic field $\mu_0 H_{C2}(0)$ = 116–158 T is more than two times larger than the calculated one (~60 T). Electrical transport measurements show that the critical current density $J_C$ in our samples may exceed 3500 A/mm$^2$ at 0 K, which opens remarkable prospects for $YH_6$ in superconducting electronics.

Anharmonic effects in $Im\bar{3}m$-$YH_6$ lead to a decrease in the EPC coefficient from 2.24 to 1.71 and lowering the critical temperature by 25 K. An anomalously large impact of the Coulomb repulsion was found in yttrium hexahydride within both the Migdal–Eliashberg and the SCDFT approaches. The calculated $T_C$ agreed with the experimental critical temperature and the isotope coefficient only when the Coulomb pseudopotential $\mu^*$(6 Ry) was equal to 0.19–0.22 and anharmonicity is included in the calculations. The parameter-free SCDFT calculations for $YH_6$ give substantially lower $T_C$ = 160 K which implies importance of effects missing in the conventional Migdal-Eliashberg theory.

# Author Contributions

[&] These authors contributed equally to this work.

D.V.S., I.A.T, A.G.I., A.G.G., V.B.P., E.G., V.V.S. performed the experiment. I.A.T. deposited electrodes and performed the superconductivity measurements. A.V.S., O.A.S., V.M.P. performed the magnetotransport experiments in high magnetic fields and participated in data processing and discussions. A.G.K. and A.R.O. prepared the theoretical analysis. A.B., I.E., R.B., M.C., F.M., L.M. calculated the anharmonic phonon density of states and Eliashberg function. R.A. performed the calculations of $T_C$ within the SCDFT formalism. D.V.S., A.G.K., A.R.O. wrote the manuscript. All authors provided critical feedback and helped shape the research, analysis, and manuscript.

# Acknowledgements

The work on the high-pressure experiments was supported by the Ministry of Science and Higher Education of the Russian Federation within the state assignment of the FSRC “Crystallography and Photonics” of RAS and by the Russian Science Foundation (Project No. 19-12-00414). A.G.K. thanks the Russian Foundation for Basic Research (project № 19-03-00100) and the Foundation for Assistance to Small Innovative Enterprises (grant UMNIK № 13408GU/2018) for the financial support of this work. A.R.O., D.V.S. and A.G.K. thank the Russian Science Foundation (grant 19-72-30043). The authors express their gratitude to the staff of the BL10XU (High Pressure Research) station of SPring-8 synchrotron research facilities, especially to Saori Kawaguchi (JASRI) for the tremendous assistance in the use of the station’s equipment before and after the experiment. Portions of this work were performed at GeoSoilEnviroCARS (The University of Chicago, Sector 13), Advanced Photon Source (APS), Argonne National Laboratory. GeoSoilEnviroCARS is supported by the National Science Foundation — Earth Sciences (EAR — 1634415) and Department of Energy — GeoSciences (DE-FG02-94ER14466). Use of the GSECARS Raman Lab System was supported by the NSF MRI Proposal (EAR-1531583). This research used the resources of the Advanced Photon Source, a U.S. Department of Energy (DOE) Office of Science User Facility operated for the DOE Office of Science by Argonne National Laboratory under contract No. DE-AC02-06CH11357. I.E.

and R.B. acknowledge support from the European Research Council (ERC) under the European Union's Horizon 2020 research and innovation programme (grant agreement no. 802533).

## References

1. Liu, H., Naumov, I. I., Hoffmann, R., Ashcroft, N. W. & Hemley, R. J. Potential high-Tc superconducting lanthanum and yttrium hydrides at high pressure. *PNAS* **114**, 6990–6995 (2017).
2. Drozdov, A. P. *et al.* Superconductivity at 250 K in lanthanum hydride under high pressures. *Nature* **569**, 528 (2019).
3. Somayazulu, M. *et al.* Evidence for Superconductivity above 260 K in Lanthanum Superhydride at Megabar Pressures. *Phys. Rev. Lett.* **122**, 027001 (2019).
4. Struzhkin, V. *et al.* Superconductivity in La and Y hydrides: Remaining questions to experiment and theory. *Matter and Radiation at Extremes* **5**, 028201 (2020).
5. APS -APS March Meeting 2020 - Event - Superconductivity in Yttrium and Thorium Polyhydrides: a Route to Industrial Applications. in *Bulletin of the American Physical Society* vol. Volume 65, Number 1 (American Physical Society).
6. Semenok, D. V. *et al.* Superconductivity at 161 K in thorium hydride ThH10: Synthesis and properties. *Mat. Today* **33**, 36–44 (2020).
7. Kruglov, I. A. *et al.* Uranium polyhydrides at moderate pressures: prediction, synthesis, and expected superconductivity. *Sci. Adv.* **4**, eaat9776 (2018).
8. Salke, N. P. *et al.* Synthesis of clathrate cerium superhydride CeH9 at 80 GPa with anomalously short H-H distance. *arXiv:1805.02060* https://arxiv.org/ftp/arxiv/papers/1805/1805.02060.pdf (2018).
9. Zhou, D. *et al.* Superconducting praseodymium superhydrides. *Science Advances* **6**, eaax6849 (2020).
10. Zhou, D. *et al.* High-Pressure Synthesis of Magnetic Neodymium Polyhydrides. *J. Am. Chem. Soc.* **142**, 2803–2811 (2020).
11. Li, Y. *et al.* Pressure-stabilized superconductive yttrium hydrides. *Scientific Reports* **5**, 09948 (2015).
12. Grishakov, K. S., Degtyarenko, N. N. & Mazur, E. A. Electron, Phonon, and Superconducting Properties of Yttrium and Sulfur Hydrides under High Pressures. *J. Exp. Theor. Phys.* **128**, 105–114 (2019).
13. Heil, C., di Cataldo, S., Bachelet, G. B. & Boeri, L. Superconductivity in sodalite-like yttrium hydride clathrates. *Phys. Rev. B* **99**, 220502 (2019).
14. Oganov, A. R. & Glass, C. W. Crystal structure prediction using ab initio evolutionary techniques: Principles and applications. *J. Chem. Phys.* **124**, 244704 (2006).
15. Oganov, A. R., Lyakhov, A. O. & Valle, M. How Evolutionary Crystal Structure Prediction Works—and Why. *Acc. Chem. Res.* **44**, 227–237 (2011).
16. Lyakhov, A. O., Oganov, A. R., Stokes, H. T. & Zhu, Q. New developments in evolutionary structure prediction algorithm USPEX. *Computer Physics Communications* **184**, 1172–1182 (2013).
17. Oganov, A. R., Ma, Y., Lyakhov, A. O., Valle, M. & Gatti, C. Evolutionary crystal structure prediction as a method for the discovery of minerals and materials. *Rev. Mineral Geochem* 271–298 (2010).
18. Kong, P. P. *et al.* Superconductivity up to 243 K in yttrium hydrides under high pressure. *arXiv:1909.10482 [cond-mat]* (2019).
19. APS -APS March Meeting 2020 - Event - Superconductivity at 262 K in Yttrium Superhydride at High Pressures. in *Bulletin of the American Physical Society* (American Physical Society).
20. Werthamer, N. R., Helfand, E. & Hohenberg, P. C. Temperature and Purity Dependence of the Superconducting Critical Field, ${H}_{c2}$. III. Electron Spin and Spin-Orbit Effects. *Phys. Rev.* **147**, 295–302 (1966).

21. Ginzburg, V. L. & Landau, L. D. On the Theory of superconductivity. *Zh.Eksp.Teor.Fiz.* **20**, 1064–1082 (1950).
22. Baumgartner, T. *et al.* Effects of neutron irradiation on pinning force scaling in state-of-the-art Nb3Sn wires. *Supercond. Sci. Technol.* **27**, 015005 (2013).
23. Carbotte, J. P. Properties of boson-exchange superconductors. *Rev. Mod. Phys.* **62**, 1027–1157 (1990).
24. Dew-Hughes, D. Flux pinning mechanisms in type II superconductors. *The Philosophical Magazine: A Journal of Theoretical Experimental and Applied Physics* **30**, 293–305 (1974).
25. Griessen, R. *et al.* Evidence for mean free path fluctuation induced pinning in YBa2Cu3O7 and YBa2Cu4O8 films. *Phys. Rev. Lett.* **72**, 1910–1913 (1994).
26. Schwartz, J. *et al.* Status of high temperature superconductor based magnets and the conductors they depend upon. *arXiv:1108.1634 [physics]* (2011).
27. Cooper, L. N. Bound Electron Pairs in a Degenerate Fermi Gas. *Phys. Rev.* **104**, 1189–1190 (1956).
28. Bardeen, J., Cooper, L. N. & Schrieffer, J. R. Microscopic Theory of Superconductivity. *Phys. Rev.* **106**, 162–164 (1957).
29. Bardeen, J., Cooper, L. N. & Schrieffer, J. R. Theory of Superconductivity. *Phys. Rev.* **108**, 1175–1204 (1957).
30. Migdal, A. B. Interaction between electrons and lattice vibrations in a normal metal. *JETP* **34**, 996–1001 (1958).
31. Eliashberg, G. M. Interactions between Electrons and Lattice Vibrations in a Superconductor. *JETP* **11**, 696–702 (1959).
32. Lüders, M. *et al.* Ab initio theory of superconductivity. I. Density functional formalism and approximate functionals. *Phys. Rev. B* **72**, 024545 (2005).
33. Marques, M. A. L. *et al.* Ab initio theory of superconductivity. II. Application to elemental metals. *Phys. Rev. B* **72**, 024546 (2005).
34. Errea, I., Calandra, M. & Mauri, F. Anharmonic free energies and phonon dispersions from the stochastic self-consistent harmonic approximation: Application to platinum and palladium hydrides. *Phys. Rev. B* **89**, 064302 (2014).
35. Bianco, R., Errea, I., Paulatto, L., Calandra, M. & Mauri, F. Second-order structural phase transitions, free energy curvature, and temperature-dependent anharmonic phonons in the self-consistent harmonic approximation: Theory and stochastic implementation. *Phys. Rev. B* **96**, 014111 (2017).
36. Monacelli, L., Errea, I., Calandra, M. & Mauri, F. Pressure and stress tensor of complex anharmonic crystals within the stochastic self-consistent harmonic approximation. *Phys. Rev. B* **98**, 024106 (2018).
37. Talantsev, E. F. Characterization of near-room-temperature superconductivity in yttrium superhydrides. *arXiv:1912.10941 [cond-mat]* (2019).
38. Uemura, Y. J. Dynamic superconductivity responses in photoexcited optical conductivity and Nernst effect. *Phys. Rev. Materials* **3**, 104801 (2019).
39. Akashi, R., Nakamura, K., Arita, R. & Imada, M. High-temperature superconductivity in layered nitrides beta-LixMNCl (M = Ti, Zr, Hf): Insights from density functional theory for superconductors. *Phys. Rev. B* **86**, 054513 (2012).
40. Kruglov, I. A. *et al.* Superconductivity of LaH10 and LaH16 polyhydrides. *Phys. Rev. B.* **101**, 024508 (2020).
41. Sanna, A. *et al.* Ab initio Eliashberg Theory: Making Genuine Predictions of Superconducting Features. *J. Phys. Soc. Jpn.* **87**, 041012 (2018).
42. Margine, E. R. & Giustino, F. Anisotropic Migdal-Eliashberg theory using Wannier functions. *Phys. Rev. B* **87**, 024505 (2013).

43. Errea, I. *et al.* Quantum crystal structure in the 250-kelvin superconducting lanthanum hydride. *Nature* **578**, 66–69 (2020).
44. Choi, H. J., Roundy, D., Sun, H., Cohen, M. L. & Louie, S. G. First-principles calculation of the superconducting transition in MgB2 within the anisotropic Eliashberg formalism. *Phys. Rev. B* **66**, 020513 (2002).
45. Sano, W., Koretsune, T., Tadano, T., Akashi, R. & Arita, R. Effect of Van Hove singularities on high-Tc superconductivity in H3S. *Phys. Rev. B* **93**, 094525 (2016).
46. Yussouff, M., Rao, B. K. & Jena, P. Reverse isotope effect on the superconductivity of PdH, PdD, and PdT. *Solid State Communications* **94**, 549–553 (1995).
47. Kvashnin, A. G., Semenok, D. V., Kruglov, I. A., Wrona, I. A. & Oganov, A. R. High-Temperature Superconductivity in Th-H System at Pressure Conditions. *ACS Applied Materials & Interfaces* **10**, 43809–43816 (2018).

# SUPPLEMENTARY INFORMATION

# Anomalous high-temperature superconductivity in $YH_6$

Ivan A. Troyan,[1,&] Dmitrii V. Semenok,[2,&,*] Alexander G. Kvashnin,[2,&,*] Andrey V. Sadakov,[3] Oleg A. Sobolevskiy,[3] Vladimir M. Pudalov,[3] Anna G. Ivanova,[1] Vitali B. Prakapenka,[4] Eran Greenberg,[4] Alexander G. Gavriliuk,[1,5] Viktor V. Struzhkin,[6] Aitor Bergara,[10,11,12] Ion Errea,[11,12,13] Raffaello Bianco,[11] Matteo Calandra,[14,15,16] Francesco Mauri,[16,17] Lorenzo Monacelli,[16,17] Ryosuke Akashi,[18] and Artem R. Oganov[2,8,9,*]

[1] Shubnikov Institute of Crystallography, Federal Scientific Research Center Crystallography and Photonics, Russian Academy of Sciences, Moscow 119333, 59 Leninskii Prospekt, Russia
[2] Skolkovo Institute of Science and Technology, Skolkovo Innovation Center, 3 Nobel Street, Moscow 121205, Russia
[3] P. N. Lebedev Physical Institute, Russian Academy of Sciences, Moscow 119991, Russia
[4] Center for Advanced Radiation Sources, The University of Chicago, 5640 South Ellis Avenue, Chicago, Illinois 60637, USA
[5] IC RAS Institute for Nuclear Research, Russian Academy of Sciences, Moscow, 117312 Russia
[6] Center for High Pressure Science and Technology Advanced Research, Shanghai 201203, China
[8] Moscow Institute of Physics and Technology, 9 Institutsky Lane, Dolgoprudny 141700, Russia
[9] International Center for Materials Discovery, Northwestern Polytechnical University, Xi'an 710072, China
[10] Departamento de Física de la Materia Condensada, University of the Basque Country (UPV/EHU), 48080 Bilbao, Basque Country, Spain
[11] Centro de Física de Materiales CFM, CSIC-UPV/EHU, Paseo Manuel de Lardizabal 5, 20018 Donostia, Basque Country, Spain
[12] Donostia International Physics Center (DIPC), Manuel Lardizabal pasealekua 4, 20018 Donostia, Basque Country, Spain
[13] Fisika Aplikatua 1 Saila, University of the Basque Country (UPV/EHU), Europa Plaza 1, 20018 Donostia, Basque Country, Spain
[14] Departimento di Fisica, Universit di Trento, Via Sommarive 14, 38123 Povo, Italy
[15] Sorbonne Université, CNRS, Institut des Nanosciences de Paris, UMR7588, F-75252 Paris, France
[16] Graphene Labs, Fondazione Istituto Italiano di Tecnologia, Via Morego, I-16163 Genova, Italy
[17] Dipartimento di Fisica, Università di Roma Sapienza, Piazzale Aldo Moro 5, I-00185 Roma, Italy
[18] University of Tokyo, 7-3-1 Hongo, Bunkyo, Tokyo 113-8654, Japan

## Table of contents

# Methods

## Experiment

To perform this experimental study, three diamond anvil cells (DACs) — K1, M1, and M3 — were loaded. The diameter of the working surface of the diamond anvils was 280 μm bevelled at an angle of 8.5° to a culet of 50 μm. The X-ray diffraction (XRD) patterns of all samples in the DACs were recorded at the GSECARS synchrotron beamline at the Advanced Photon Source (APS), Argonne, U.S. An X-ray beam with an energy of 42 and 37 keV and wavelength λ = 0.295 and 0.334 Å was focused to 2.5×3.5 μm. A Pilatus 1M CdTe detector was placed at a distance of ~200 mm from the sample. The exposure time was 20–60 s. $LaB_6$ standard was used for the detector geometry calibration. The XRD data were analysed and integrated using Dioptas software package (version 0.4).[1] The full profile analysis of the diffraction patterns and the calculation of the unit cell parameters were performed in JANA2006 computing system[2] using the Le Bail method.[3]

The heating of the sample was performed at GSECARS by ~$10^5$ pulses of a Nd:YAG infrared laser with a wavelength λ = 1.064 μm, the duration of each pulse was 1 μs.[4] The temperature measurements were carried out using the gray body radiation fit within the Planck function at the laser heating system of the GSECARS beamline of the APS. The applied pressure was measured by the edge position of the Raman signal of diamond[5] using Acton SP2500 spectrometer with PIXIS:100 spectroscopic-format CCD.[6] The pressure in the DACs was determined by the Raman signal of diamond.[7]

Deuterated ammonium borane (d-AB) was synthesized from $NaBD_4$ (98 % D, SigmaAldrich) via the reaction with ammonium formate $HCOONH_4$ in tetrahydrofuran followed by isotopic substitution (H→D) in $D_2O$ [8]. After removing of solvents and vacuum drying, the obtained $ND_3BD_3$ was analyzed by $^1$H NMR and Raman spectroscopy. Deuterium content in the product was found to be 92 %. Partially substituted $NH_3BD_3$ and $ND_3BH_3$ may be synthesized in a similar way for later use as a source of HD.

Magnetotransport measurements were performed on samples with at least two hydride phases and, therefore, the voltage contacts of the Van der Pauw scheme might be connected to low-$T_C$ phase. As a result the superconducting transition in $YH_6$ can be observed as an upward feature of the R(T,H) curves due to the shunting effect in the fine-grained samples.

For the Dew-Hughes model [9] of the pinning force $f \sim h^p(1-h)^q$, the parameters should be p=0.5, q=2, $h_{max}$=0.2, which is close to the fit of the experimental data (Fig. S13a). Depinning critical current for this type of pinning can be described within the single vortices model, where vortices are pinned on randomly distributed weak pinning centers via spatial fluctuations of the charge carrier mean free path, or in other words "dl-pinning".

**Table S1.** Experimental parameters of the DACs.

| DAC | Pressure, GPa | Gasket | Sample size, μm | Composition/load |
|---|---|---|---|---|
| K1 | 166 | $CaF_2$/epoxy | 10 | Y/$NH_3BH_3$ |
| M1 | 165 | $CaF_2$/epoxy | 12 | Y/$NH_3BH_3$ |
| M3 | 172 | $CaF_2$/epoxy | 9 | Y/$NH_3BH_3$ |
| D1 | 172 | $CaF_2$/epoxy | 15 | Y/$ND_3BD_3$ |

## Theory

The computational predictions of the thermodynamic stability of the Y–H phase at 150, 200, and 250 GPa were carried out using the variable-composition evolutionary algorithm USPEX.[10–12] The first generation consisting of 120 structures was produced using random symmetric[12] and random topology[13] generators, whereas all subsequent generations contained 20% of random structures and 80% of those created using heredity, softmutation, and transmutation operators. The evolutionary searches were combined with structure relaxations using the density functional theory (DFT)[14,15] within the Perdew–Burke–Ernzerhof functional (generalized gradient approximation)[16] and the projector augmented wave method[17,18] as implemented in the VASP code.[19–21] The kinetic energy cutoff for plane waves was 600 eV. The Brillouin zone was sampled using the Γ-centred *k*-points meshes with a resolution of $2\pi\times0.05$ Å$^{-1}$. The methodology is similar to those used in our previous works.[22,23]

The equations of state of the discovered $YH_4$, $YH_6$, and $YH_7$ phases were calculated using the same methods with the plane wave kinetic energy cutoff set to 700 eV. We also calculated the phonon densities of states of the studied materials using the finite displacements method (VASP and PHONOPY).[24,25]

The calculations of superconducting $T_C$ were carried out using QUANTUM ESPRESSO (QE) package.[26,27] The phonon frequencies and electron–phonon coupling (EPC) coefficients were computed using the density functional perturbation theory,[28] employing the plane-wave pseudopotential method and Perdew–Burke–Ernzerhof exchange-correlation functional.[16] In our ab initio calculations of the electron-phonon coupling (EPC) coefficient λ, the first Brillouin zone was sampled using a 3×3×3 or 4×4×4 *q*-points mesh and a denser 16×16×16 *k*-points mesh (with the Gaussian smearing and σ = 0.005 Ry, which approximates the zero-width limits in the calculation of λ) for $YH_7$ and $YH_4$.

The calculations of the Eliashberg function of $Im\bar{3}m$-$YH_6$ were performed using 6×6×6 *q*-points and 60×60×60 *k*-points meshes. $T_C$ was calculated by solving the isotropic Eliashberg equations[29] using the iterative self-consistent method for the imaginary part of the order parameter $\Delta(T, \omega)$ (superconducting gap) and the renormalization wave function $Z(T, \omega)$ assuming that the Coulomb repulsion between the electrons can be parametrized with the $\mu^*$. More approximate estimates of $T_C$ were made using the Allen–Dynes formula.[30]

We also calculated the superconducting transition temperature $T_C$ of $YH_6$ by solving the gap equation in the density functional theory for superconductors (SCDFT):[31,32]

$$\Delta_{nk}(T) = -Z_{nk}(T)\Delta_{nk}(T) - \frac{1}{2}\sum K_{nkn'k'}(T)\frac{\tanh \beta\xi_{n'k'}}{\xi_{n'k'}}\Delta_{n'k'}(T). \quad \text{(S1)}$$

Solving this equation for different temperatures $\beta = 1/T$, we see that the order parameter $\Delta_{n'k'}$ has a nonzero solution below any threshold temperature, which is identified as $T_c$. Labels *n*, *n'*, *k*, and *k'* indicate the Kohn–Sham band and crystal wave number indices, respectively. $\xi_{nk}$ is the energy eigenvalue of state *nk* from the Fermi level, as calculated using the standard Kohn–Sham equation for the normal state. The kernels of the gap equation $Z_{nk}(T)$ and $K_{nkn'k'}(T)$ represent the electron–phonon and electron–electron Coulomb interaction effects, the formulas for which (a study by Kruglov et al.[33] based on the method described in Refs. [31,32,34,35]) have been constructed so that the perturbation effects, almost the same as those in the Eliashberg equations with the Migdal approximation,[29,36–38] are included. The solution of equation S1 requires preprocessing to calculate the dielectric matrix for the screened electron–electron Coulomb interaction within the random phase approximation[39] and the electronic density of states (DOS) for the normal state. The detailed conditions for the SCDFT calculation are summarized in Table S2.

**Table S2.** Detailed conditions for calculating $T_c$ of $YH_6$ within the SCDFT approach.

| Crystal structure setting | | $(YH_6)_2$ simple cubic |
|---|---|---|
| Charge density | *k* | 12×12×12 equal mesh |
| | Interpolation | 1st order Hermite Gaussian[39] with a width of 0.020 Ry |
| Dielectric matrix ε | *k* for bands crossing $E_F$ | 15×15×15 equal mesh |
| | *k* for other bands | 5×5×5 equal mesh |
| | Number of unoccupied bands† | 82 |
| | Interpolation | Tetrahedron with the Rath–Freeman treatment[40] |
| | cutoff | 12.8 Ry |
| DOS for phononic kernels | *k* | 19×19×19 equal mesh |
| | interpolation | Tetrahedron with the Blöchl correction[17] |
| SCDFT gap function | Number of unoccupied bands‡ | 82 |
| | *k* for the electronic kernel | 5×5×5 equal mesh |
| | *k* for the KS energies | 19×19×19 equal mesh |
| | Sampling points for bands crossing $E_F$ | 6000 |
| | Sampling points for other bands | 200 |
| | Sampling error in $T_c$, % | ~2.5 |

†States up to $E_F$ + 70 eV are taken into account. ‡States up to $E_F$+70 eV are taken into account.

The anharmonic calculations, including the vibrational contribution to the enthalpy, were performed using the stochastic self-consistent harmonic approximation (SSCHA).[41] The anharmonic force constant matrices of $Im\bar{3}m$-$YH_6$ were obtained by calculating the forces in 3×3×3 supercells and combined with the DFPT electron–phonon calculations performed in a fine 12×12×12 mesh to calculate the anharmonic Eliashberg function $\alpha^2F(\omega)$.

# Crystal structures

**Table S3.** Crystal data of the predicted yttrium hydrides and pure elements at 150 GPa.

| Phase | Volume, Å$^3$/atom | Lattice parameters | Coordinates | | | |
|---|---|---|---|---|---|---|
| | | | atom | x | y | z |
| $Fddd$-Y | 12.88 | $a$ = 16.867 Å<br>$b$ = 4.564 Å<br>$c$ = 2.677 Å | Y1 | -0.187 | 0.125 | 0.125 |
| $C2/c$-H | 1.86 | $a$ = 5.332 Å | H1 | 0.261 | 0.083 | 0.254 |
| | | $b$ = 3.052 Å | H2 | 0.148 | 0.197 | 0.271 |
| | | $c$ = 4.484 Å | H3 | 0.000 | -0.159 | 0.250 |
| | | β = 142.25 º | H4 | 0.000 | -0.403 | 0.250 |
| $Fm\bar{3}m$-YH | 7.643 | $a$ = 3.939 Å | Y1 | 0.500 | 0.500 | 0.500 |
| | | | H1 | 0.000 | 0.000 | 0.000 |
| $I4/mmm$-$YH_3$ | 4.927 | $a$ = 2.995 Å | Y1 | 0.000 | 0.000 | 0.000 |
| | | $c$ = 4.392 Å | H1 | 0.000 | 0.500 | 0.250 |
| | | | H2 | 0.000 | 0.000 | 0.500 |
| $I4/mmm$-$YH_4$ | 4.072 | $a$ = 2.799 Å | Y1 | 0.000 | 0.000 | 0.000 |
| | | $c$ = 5.278 Å | H1 | 0.000 | 0.000 | -0.371 |
| | | | H2 | 0.000 | 0.500 | 0.250 |
| $Im\bar{3}m$-$YH_6$ | 3.339 | $a$ = 3.602 Å | Y1 | 0.000 | 0.000 | 0.000 |
| | | | H1 | 0.250 | 0.000 | 0.500 |
| $Imm2$-$YH_7$ | 3.261 | $a$ = 3.281 Å | Y1 | 0.000 | 0.000 | 0.018 |
| | | $b$ = 3.402 Å | H1 | 0.000 | -0.268 | -0.333 |
| | | $c$ = 4.676 Å | H2 | 0.203 | 0.000 | 0.447 |
| | | | H3 | -0.263 | 0.000 | 0.384 |
| | | | H4 | 0.000 | 0.500 | 0.299 |
| $P1$-$YH_7$ | 3.261 | $a$ = 3.298 Å | Y1 | 0.348 | 0.104 | -0.351 |
| | | $b$ = 3.306 Å | Y2 | 0.000 | -0.224 | 0.148 |
| | | $c$ = 5.520 Å | H1 | 0.301 | 0.125 | 0.261 |
| | | α = 90.057º | H2 | -0.328 | 0.440 | -0.336 |
| | | β = 91.257º | H3 | 0.493 | -0.212 | -0.045 |
| | | γ = 60.138º | H4 | 0.292 | -0.370 | 0.462 |
| | | | H5 | -0.149 | 0.316 | -0.127 |
| | | | H6 | -0.005 | -0.228 | -0.240 |
| | | | H7 | -0.388 | -0.326 | 0.370 |
| | | | H8 | -0.144 | -0.395 | 0.452 |
| | | | H9 | -0.343 | 0.446 | 0.163 |
| | | | H10 | 0.069 | 0.246 | -0.038 |
| | | | H11 | 0.487 | 0.264 | -0.041 |
| | | | H12 | 0.395 | -0.402 | -0.133 |
| | | | H13 | -0.145 | 0.092 | 0.452 |
| | | | H14 | -0.393 | 0.266 | 0.369 |
| $P6_3/mmc$-$YH_9$ | 3.051 | $a$ = 3.405 Å | Y1 | 0.333 | 0.666 | 0.750 |
| | | $c$ = 5.772 Å | H1 | -0.175 | -0.351 | -0.061 |
| | | | H2 | 0.000 | 0.000 | -0.157 |
| | | | H3 | 0.333 | 0.666 | 0.250 |
| $F\bar{4}3m$-$YH_9$ | 2.886 | $a$ = 4.869 Å | Y1 | 0.500 | 0.500 | 0.500 |
| | | | H1 | 0.135 | 0.135 | 0.135 |
| | | | H2 | -0.116 | -0.116 | -0.116 |
| | | | H3 | 0.750 | 0.0.75 | 0.0.75 |
| $Fm\bar{3}m$-$YH_{10}$ | 2.726 | $a$ = 4.931 Å | Y1 | 0.000 | 0.000 | 0.000 |
| | | | H1 | 0.378 | 0.378 | 0.378 |
| | | | H2 | 0.250 | 0.250 | 0.250 |

# Thermodynamic stability of Y–H compounds

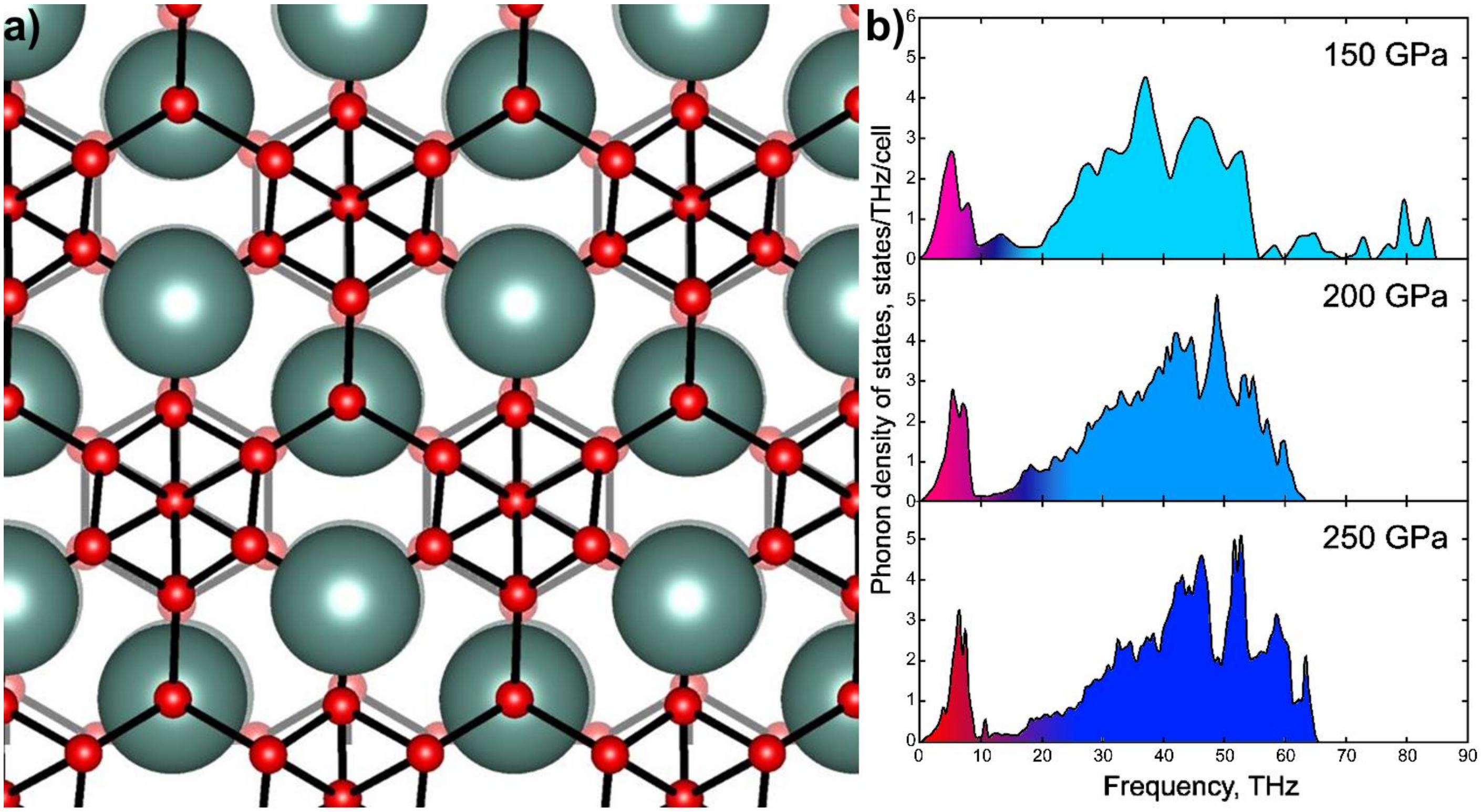

**Figure S1.** (a) Crystal structure of P$\bar{1}$-$YH_9$ (=$Y_4H_{36}$) with the disordered hydrogen sublattice in comparison with ideal symmetric *P*6$_3$/*mmc*-$YH_9$ (shaded). The hydrogen atoms are shown in red. (b) Phonon density of states of *P*$\bar{1}$-$YH_9$ at 150, 200, and 250 GPa.

**Table S4.** Crystal data of predicted pseudohexagonal yttrium hydride *P*$\bar{1}$-$YH_9$ (=$Y_4H_{36}$) at 200 GPa.

| Phase | Volume, Å$^3$/atom | Lattice parameters | Coordinates |
|---|---|---|---|
| *P*$\bar{1}$-$YH_9$ | 2.662 | *a* = 3.373 Å<br>*b* = 5.341 Å<br>*c* = 3.385 Å<br>α = 89.998°<br>β = 119.194 °<br>γ = 90.129 ° | Y1 -0.16542 -0.25018 0.16548<br>H1 0.49959 -0.25006 -0.49955<br>H2 -0.18944 -0.44497 -0.35591<br>H3 -0.18379 -0.05639 -0.35437<br>H4 0.35525 -0.05117 -0.35839<br>H5 0.34935 -0.05657 0.17690<br>H6 -0.16339 0.15138 0.16471<br>H7 0.35110 -0.44761 -0.35842<br>H8 0.34575 -0.44296 0.17626<br>H9 0.16721 -0.34794 -0.16444 |

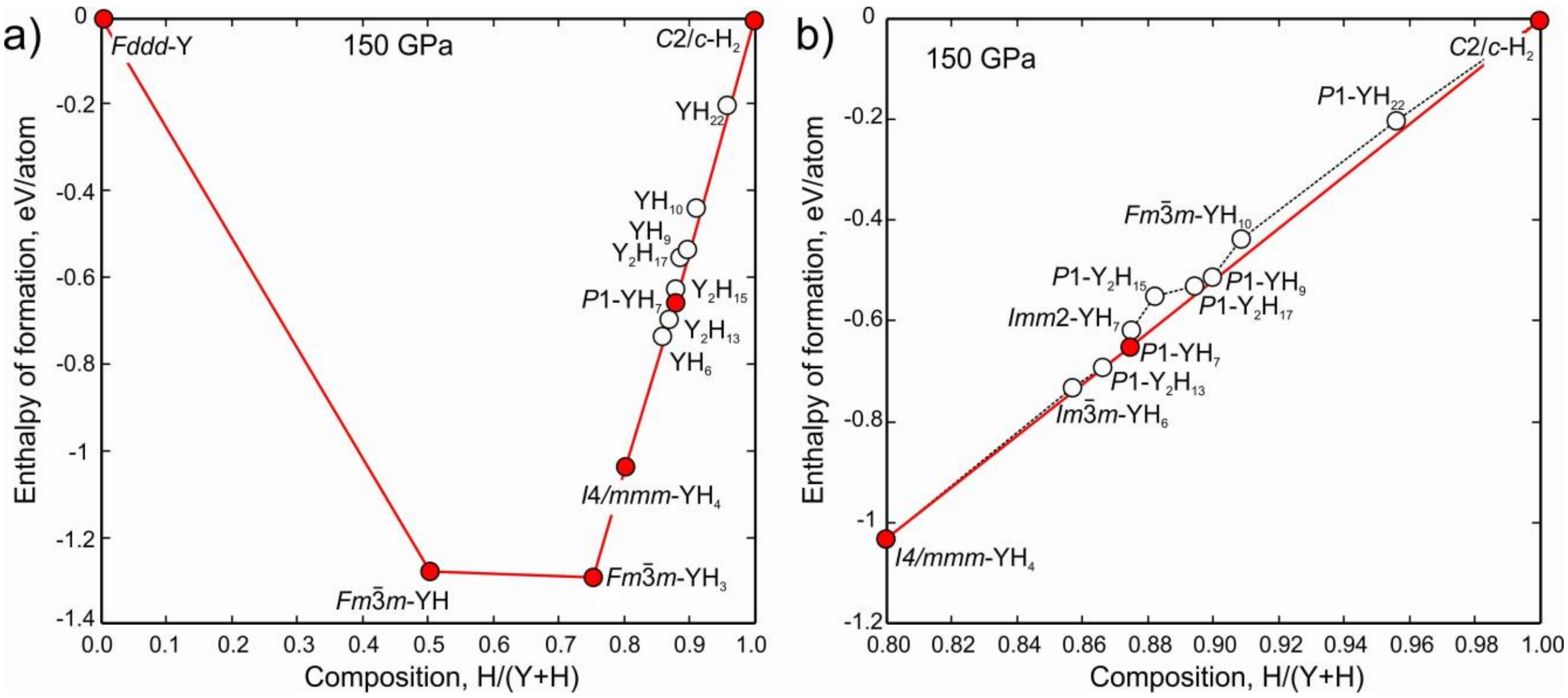

**Figure S2.** Calculated convex hull of the Y–H system with the zero-point energy (ZPE) at 150 GPa and 0 K: (a) full scale and (b) compositions from 0.8 to 1. Metastable phases are shown by white circles.

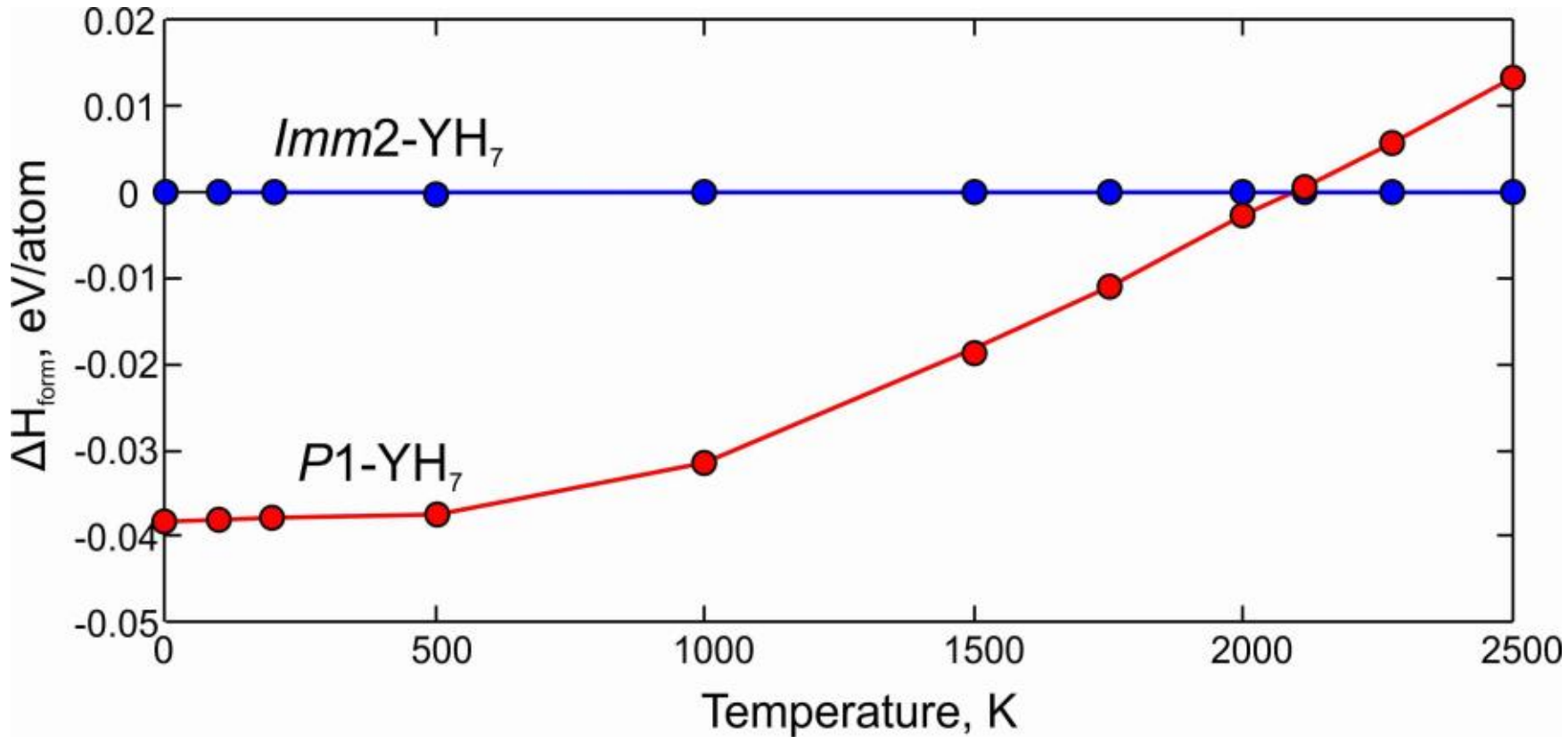

**Figure S3.** Relative enthalpy of formation of *P*1-$YH_7$ and *Imm*2-$YH_7$ at 150 GPa.

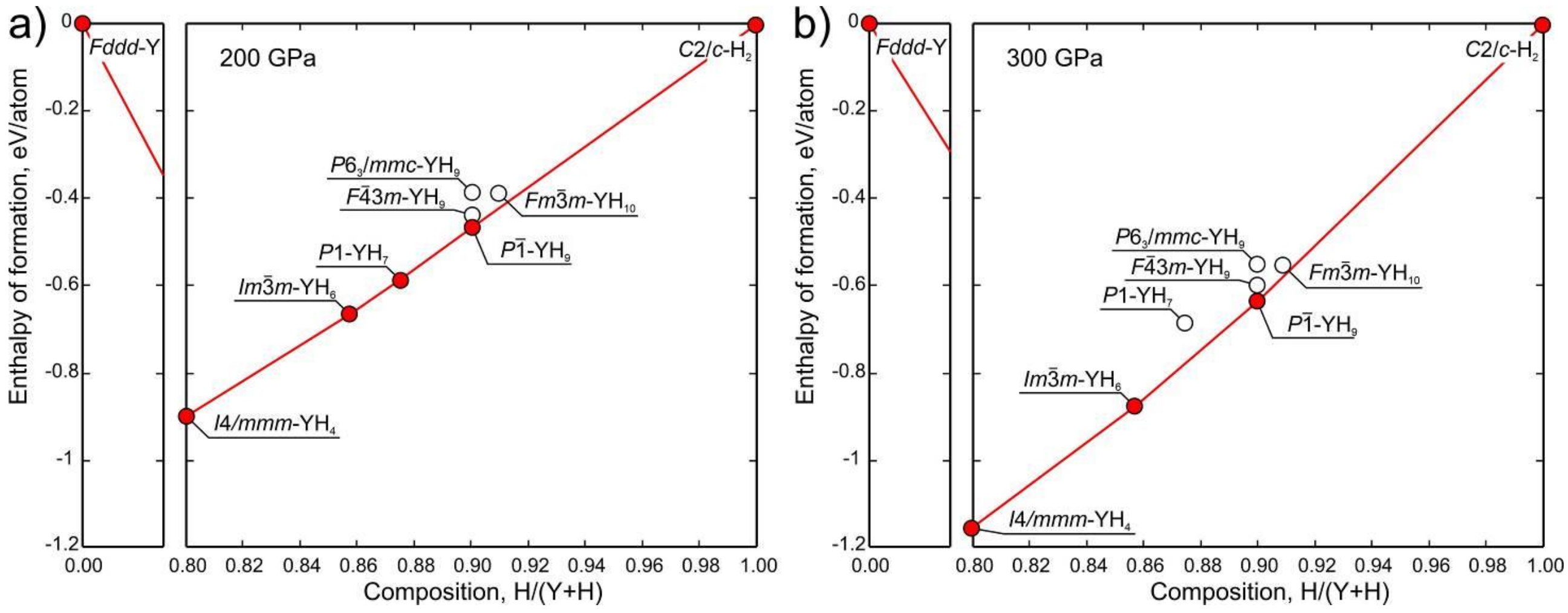

**Figure S4.** Calculated convex hull of the Y–H system with ZPE at (a) 200 and (b) 300 GPa and 0 K.

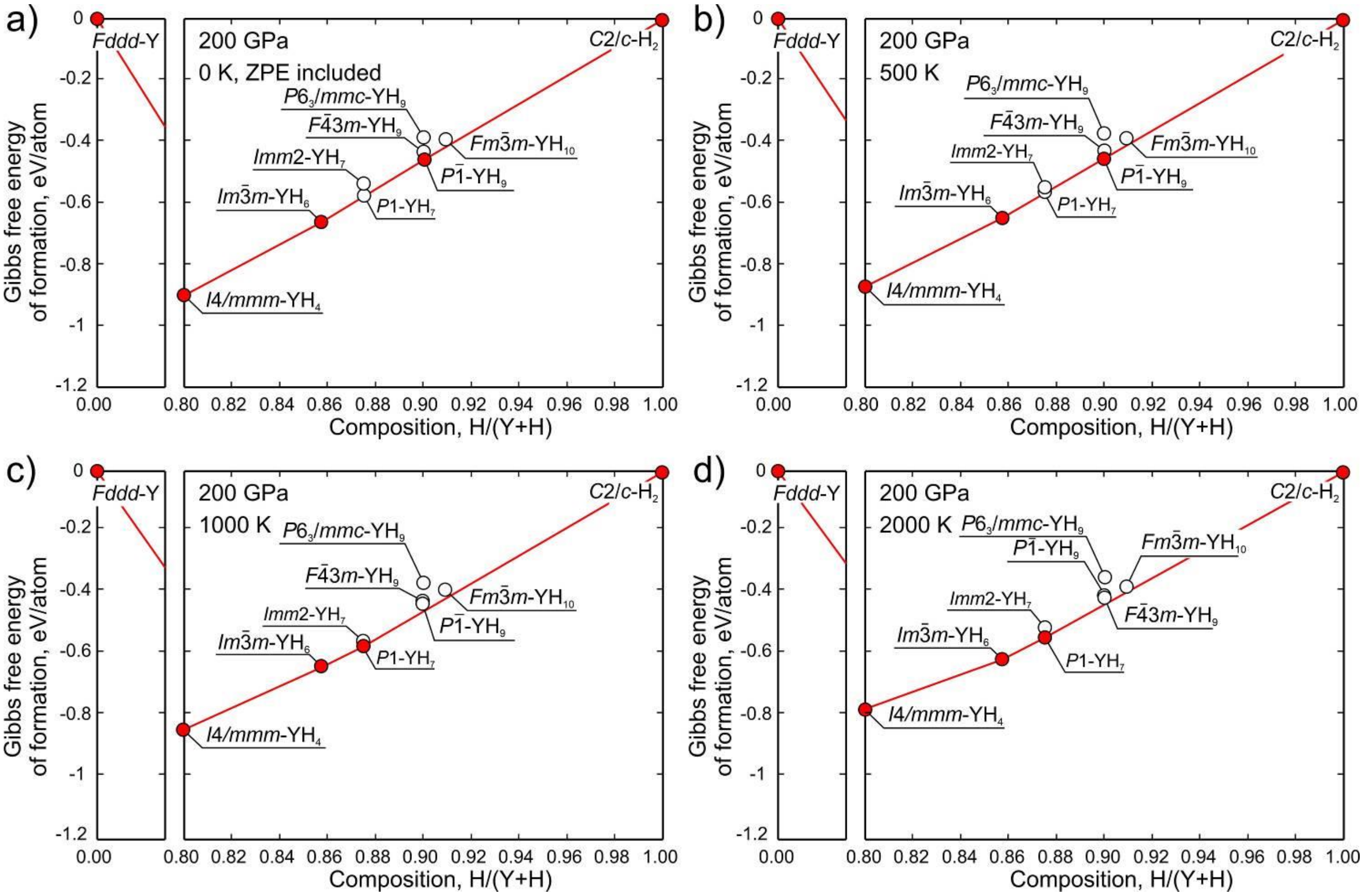

**Figure S5.** Calculated convex hulls of the Y–H system at 200 GPa and (a) 0 K, (b) 500 K, (c) 1000 K, and (d) 2000 K.

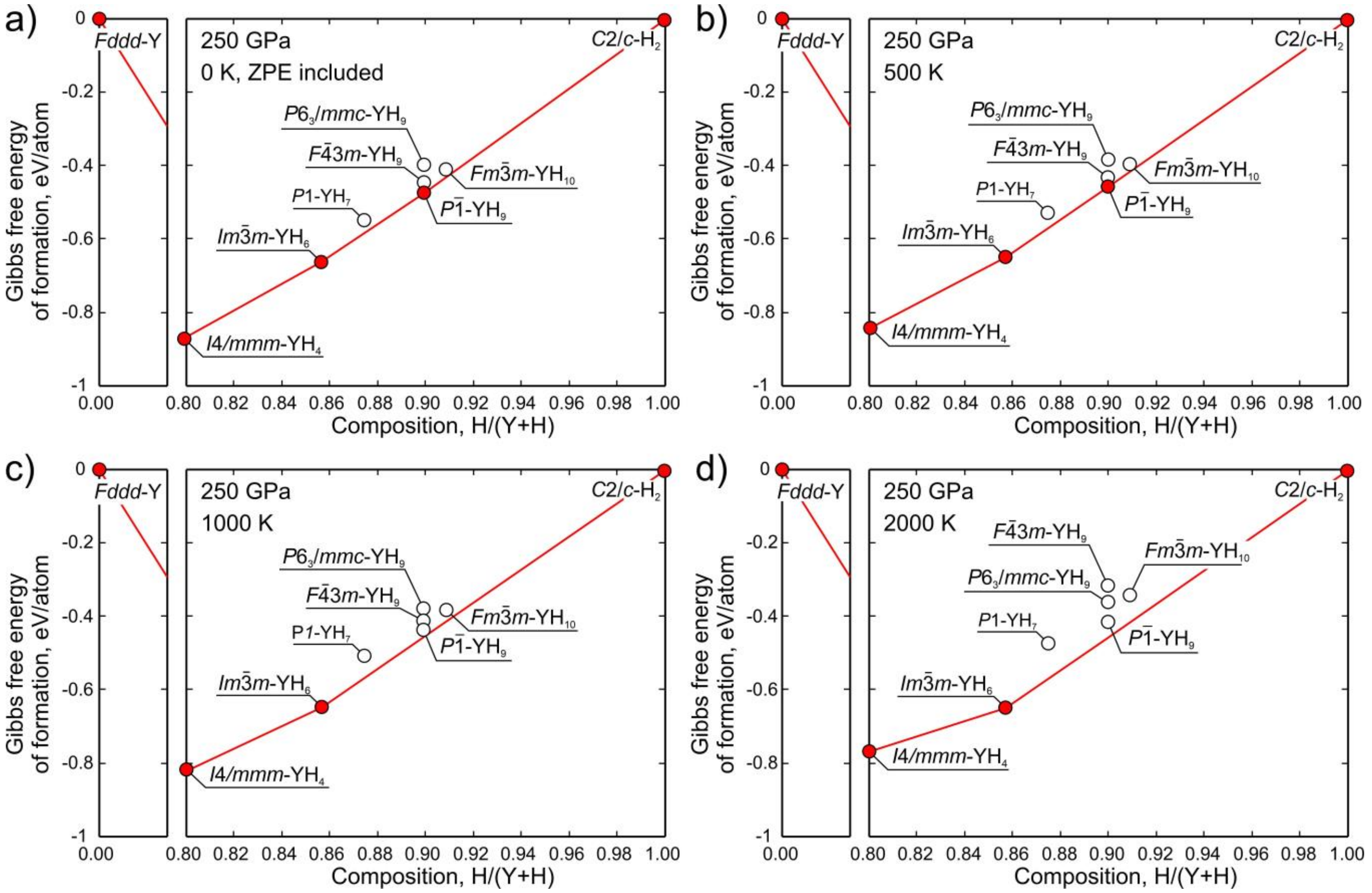

**Figure S6.** Calculated convex hulls of the Y–H system at 250 GPa and (a) 0 K, (b) 500 K, (c) 1000 K, and (d) 2000 K.

# X-ray diffraction data

**Table S5.** Experimental and predicted lattice parameters and volumes of *I*4/*mmm*-$YH_4$ ($Z$ = 2), *Imm*2-$YH_7$ ($Z$ = 2), and *P*1-$YH_7$ ($Z$ = 2).

| ***I*4/*mmm*-$YH_4$** | | | | | |
|---|---|---|---|---|---|
| ***DAC*** | Pressure, GPa | *a*, Å | *c*, Å | *V*, Å$^3$ | $V_{DFT}$, Å$^3$ |
| **K1** | 168 | 2.751(4) | 5.15(8) | 39.01 | 40.04 |
| **K3** | 173 | 2.68(9) | 5.43(3) | 39.29 | 39.70 |
| **K3** | 180 | 2.67(6) | 5.39(6) | 38.65 | 39.28 |

| ***Imm*2-$YH_7$** | | | | | | |
|---|---|---|---|---|---|---|
| ***DAC*** | Pressure, GPa | *a*, Å | *b*, Å | *c*, Å | *V*, Å$^3$ | $V_{DFT}$, Å$^3$ |
| **M1** | 166 | 3.29(4) | 3.33(6) | 4.68(7) | 51.50 | 50.85 |

| ***P*1-$YH_7$** | | | | | | | | | |
|---|---|---|---|---|---|---|---|---|---|
| ***DAC*** | Pressure, GPa | *a*, Å | *b*, Å | *c*, Å | α | β | γ | *V*, Å$^3$ | $V_{DFT}$, Å$^3$ |
| **M1** | 166 | 3.22(4) | 3.27(1) | 5.43(8) | 93 | 94.05 | 61.39 | 50.22 | 50.88 |

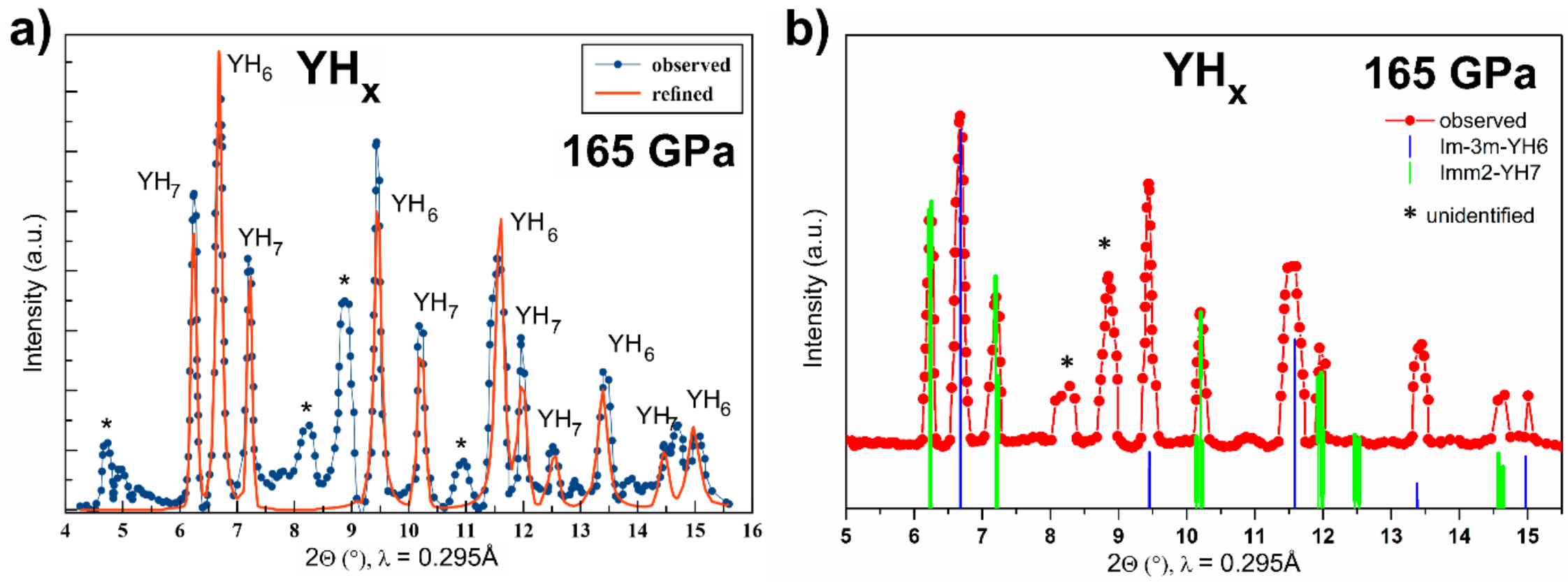

**Figure S7.** Diffraction patterns (low intensity) of the sample from DAC M1 at 165 GPa: (a) Le Bail refinement by *Im*$\bar{3}$*m*-$YH_6$ and pseudocubic *Imm*2-$YH_7$; (b) qualitative interpretation of the smoothed XRD pattern with subtracted background.

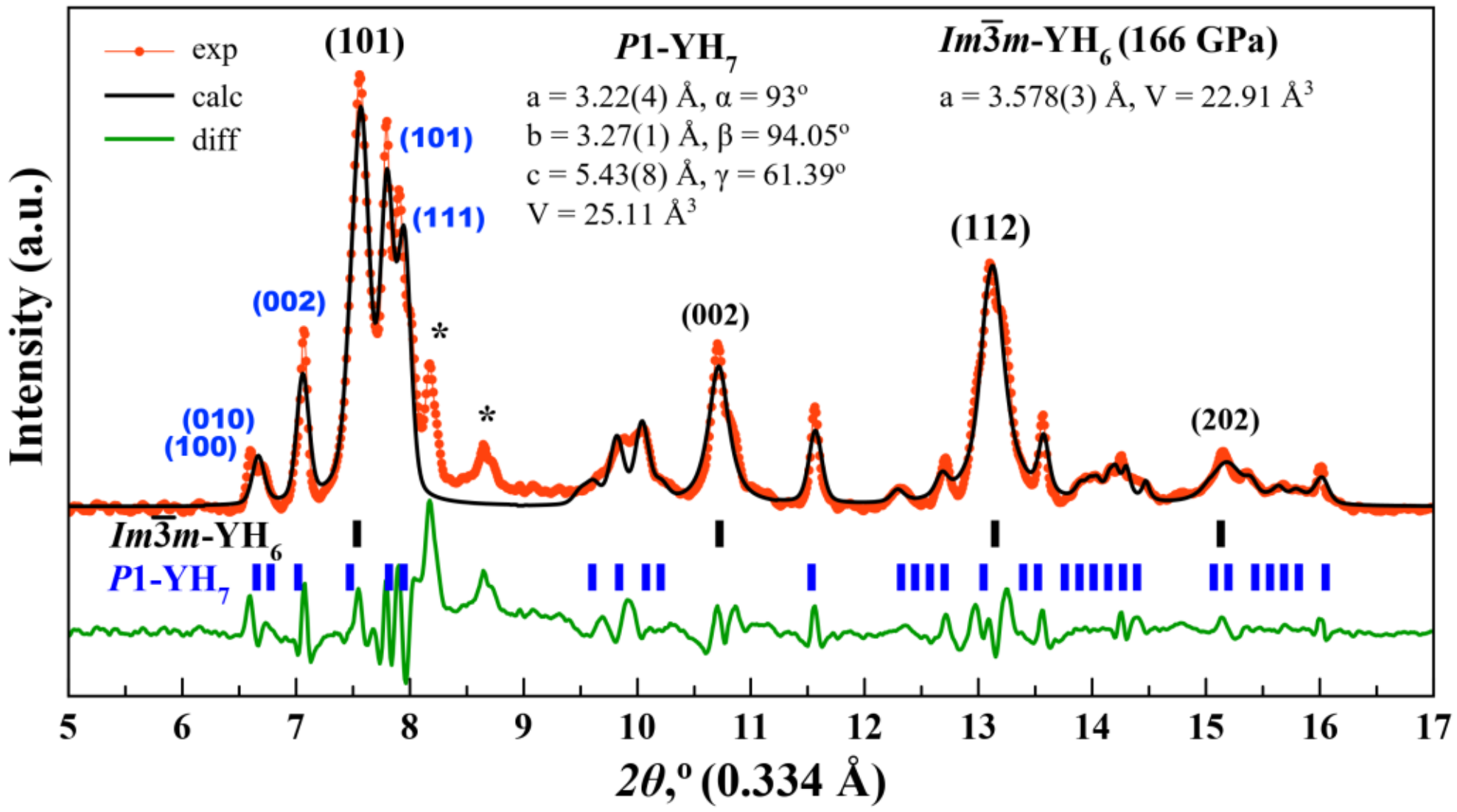

**Figure S8.** Le Bail refinement of *P*1-$YH_7$ and the experimental XRD pattern at 166 GPa (DAC K1). The experimental data, model fit for the structure, and residues are shown in red, black, and green, respectively. Unidentified reflections are marked by asterisks.

# Raman spectra

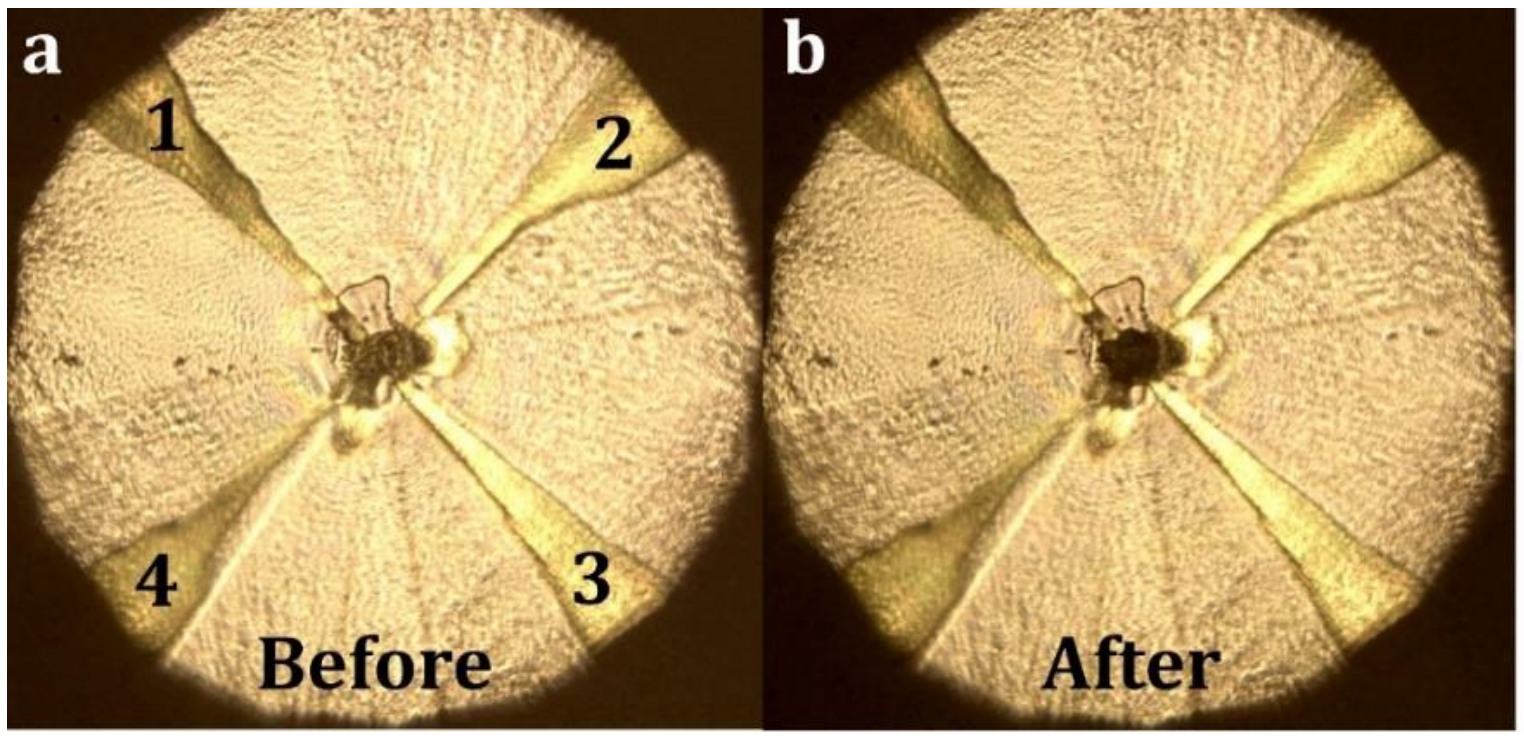

**Figure S9.** Electrode system of DAC K1 at 166 GPa (a) before and (b) after laser heating. The heating area in the centre of the culet became black.

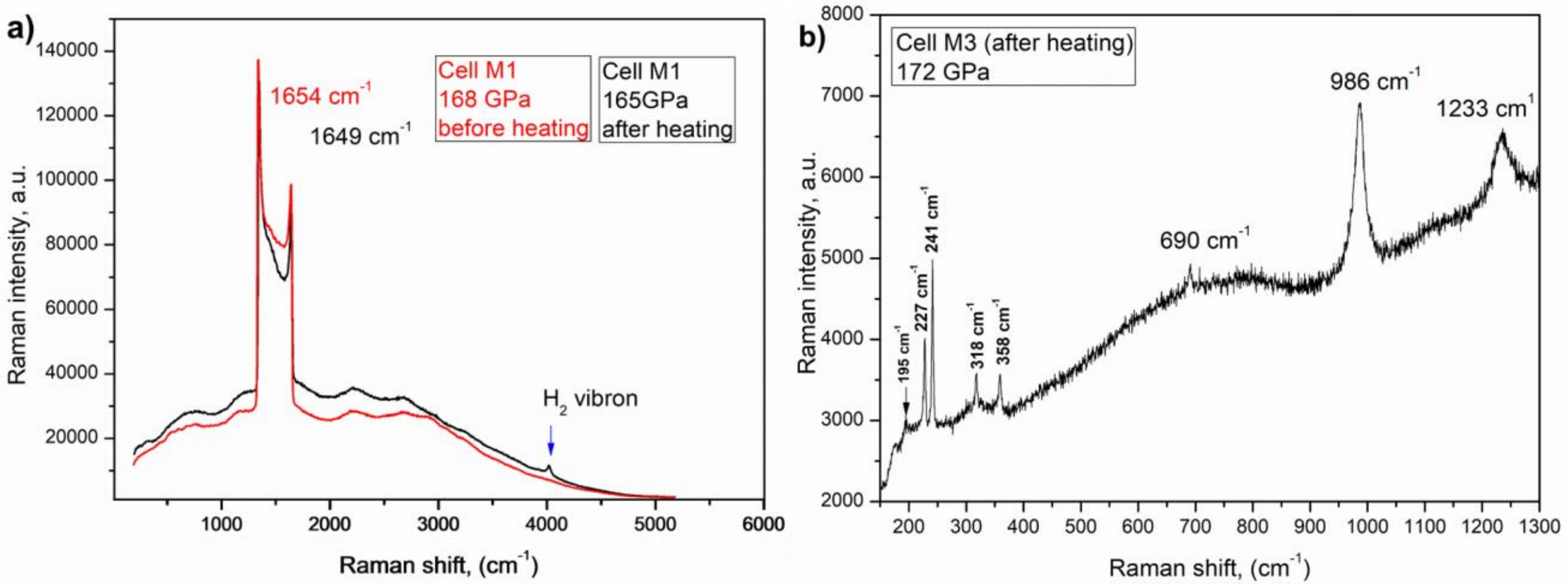

**Figure S10.** Raman spectra of samples in (a) DAC M1 and (b) DAC M3, before and after heating. Raman signals in DAC M3 may come from impurities of higher molecular yttrium hydrides ($P1$-$YH_7$-$YH_{7.5}$).

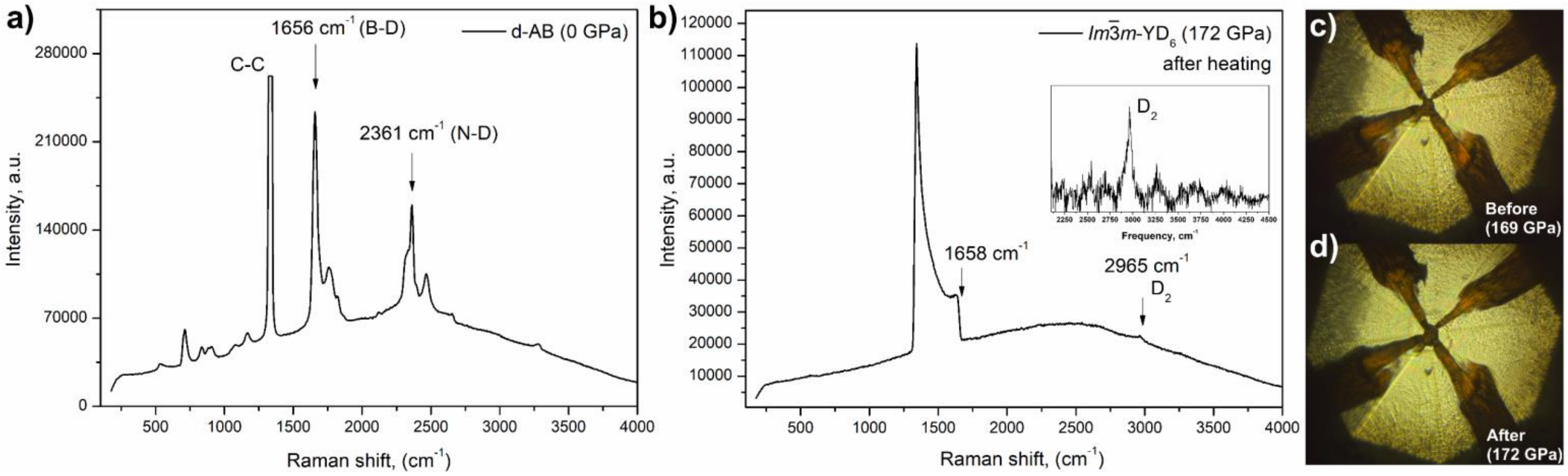

**Figure S11.** Raman spectra of samples: (a) $ND_3BD_3$ before the experiment; (b) $Im\bar{3}m$-$YD_6$ after heating (inset: Raman signal of $D_2$; (c) electrode system of the DAC before and (d) after laser heating.

# Critical parameters of superconducting state of $YH_6$

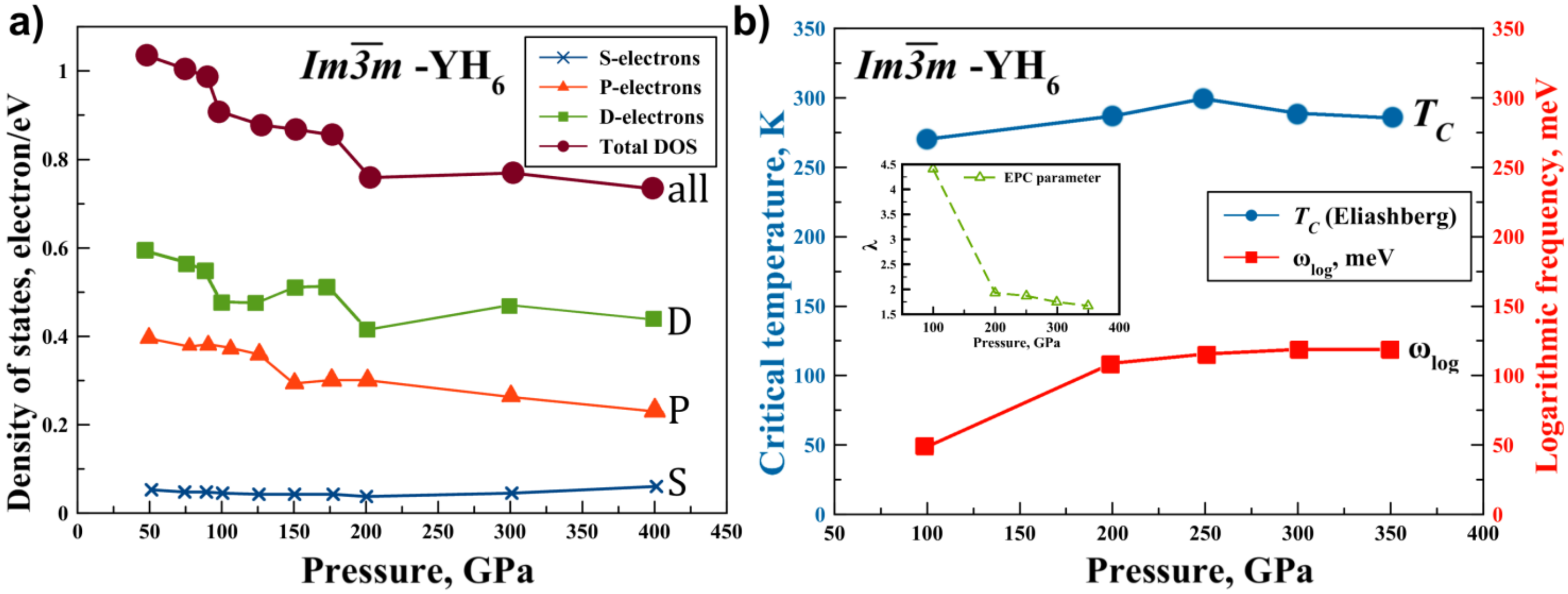

**Figure S12.** Yttrium hexahydride properties, from previous theoretical studies (see Table S6): (a) contributions of different orbitals to the density of electronic states $N(E_F)$ from Ref. [42]; (b) pressure dependence of superconducting parameters $T_c$, λ, and $\omega_{log}$ from Ref. [43].

**Table S6.** Calculated superconducting parameters of $Im\bar{3}m$-$YH_6$ from previous theoretical studies.

| Parameter | 100 GPa | 120 GPa | 200 GPa | 300 GPa |
|---|---|---|---|---|
| $N(E_F)$, states/eV/f.u. | 0.75[42] | 0.6[44] | 0.63[42] | 0.69[43] |
| λ | 3.44* | 2.93[44] | 1.93 | 1.73[43] |
| $\omega_{log}$, K | 851* | 1080,[44] 1124[45] | 1282 | 1612[43] |
| $T_C$, K | 233* (A–D) | 251–264[44] | 285[43] | 290[43] |
| Anharmonic $\Delta T_c$, K | - | –34[46] | ~ 0[43] | ~ 0[43] |

* Calculated from $\alpha^2F(\omega)$ given in Ref. [43] at μ* = 0.1.

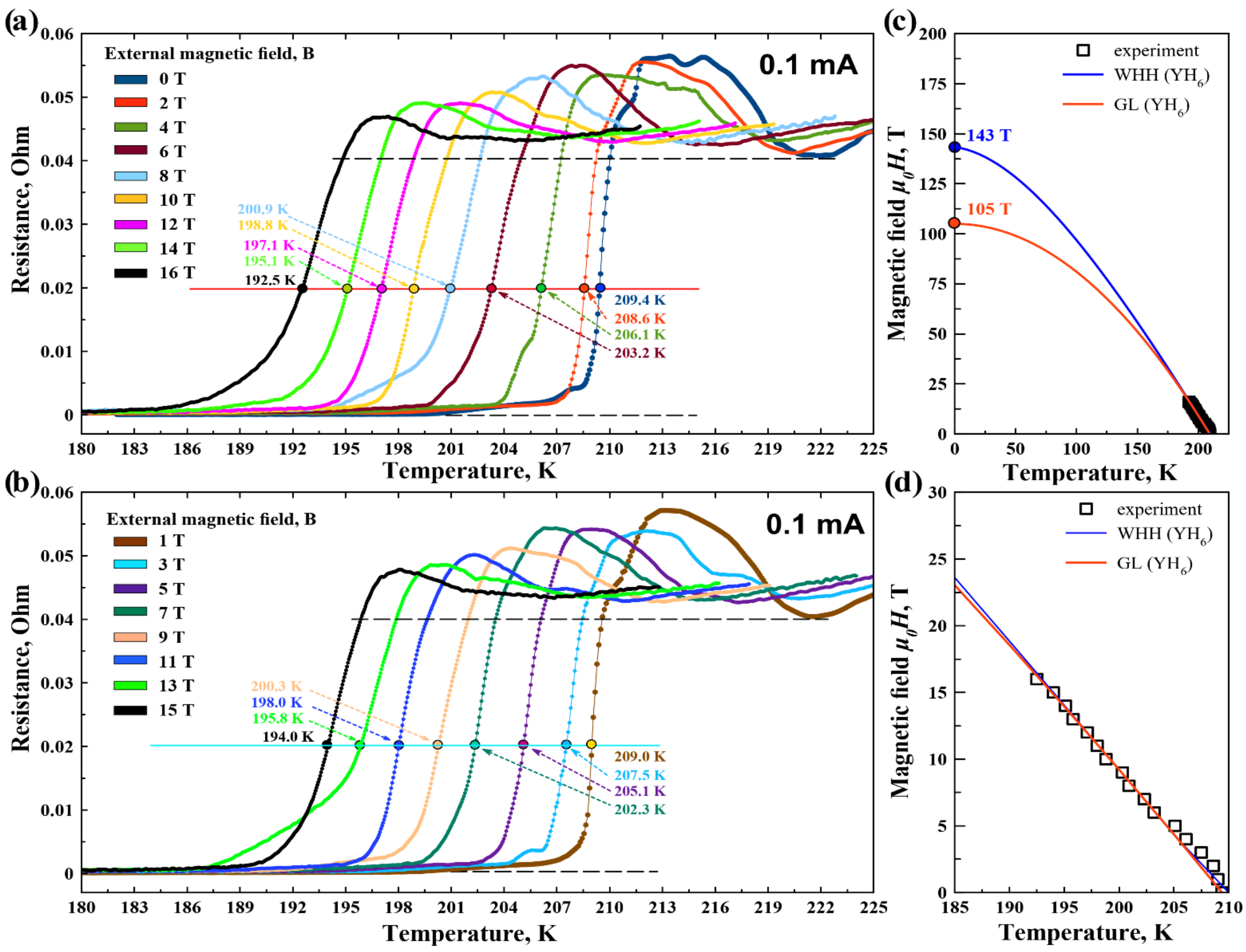

**Figure S13.** (a, b) Dependence of the electrical resistance (second step) of $Im\bar{3}m$-$YH_6$ on the external magnetic field (0–16 T) at 183 GPa: (a) even and (b) odd values of the magnetic field. Critical temperatures were determined at a 50% resistance drop; (c) upper critical magnetic field determined using different theories; (d) dependence of the critical temperature $T_C$ on the applied magnetic field.

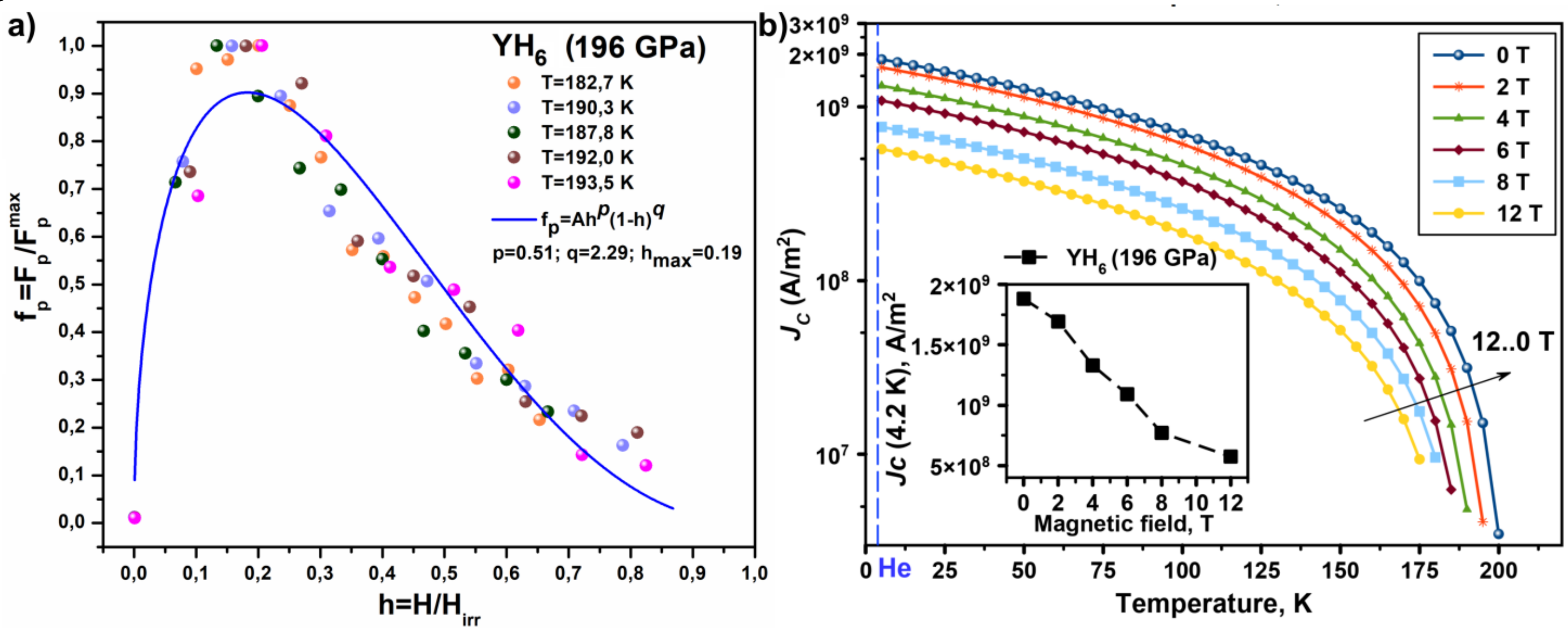

**Figure S14**. (a) Scaling of the normalized volume pinning forces ($F_p/F_p^{max}$) of the $YH_6$ sample at 196 GPa for several different temperatures versus the reduced field h = $B/B_{max}$. Experimental data is fitted by the Dew-Hughes [9] model for surface type normal pinning centers $f \sim h^p(1-h)^q$ . (b) Extrapolation of the temperature dependence of the critical current using Landau-Ginzburg model [47] $J_C = J_{c0}(1 - T/T_C)^{3/2}$; inset: dependencies of the critical current density at 4.2 K on the magnetic field.

## Eliashberg functions of yttrium hydrides

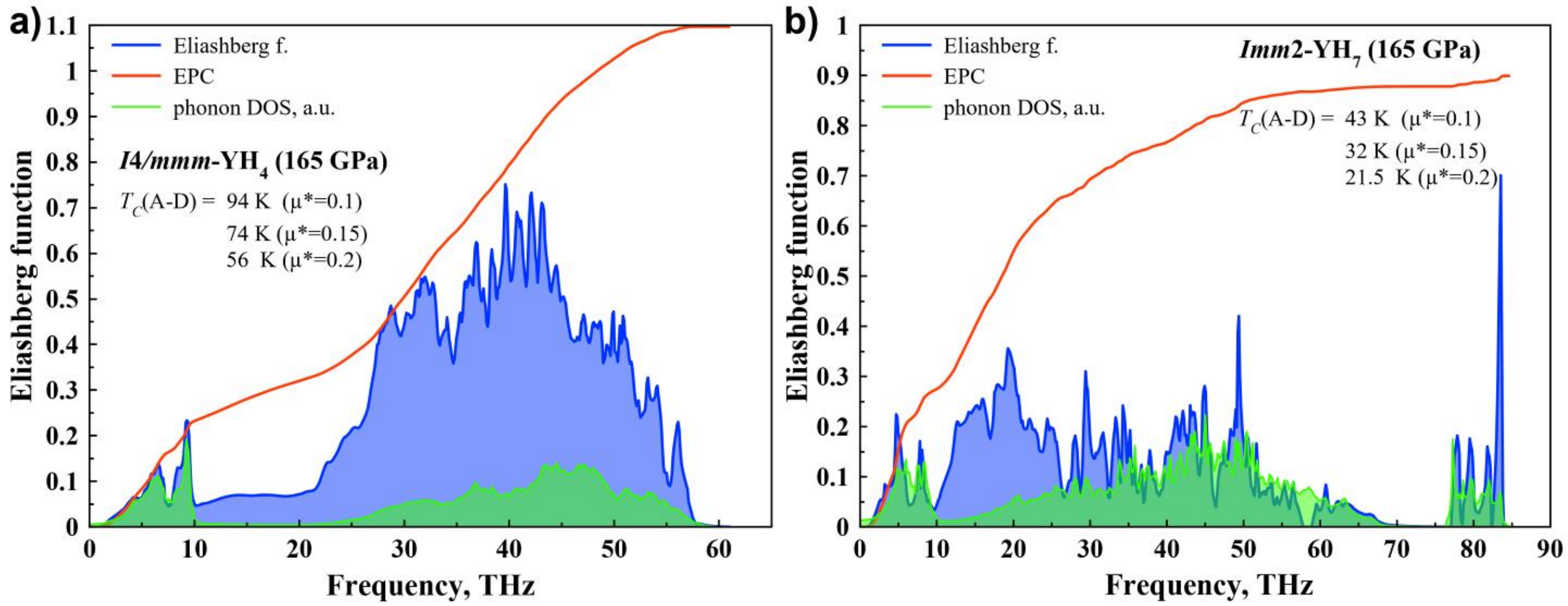

**Figure S15.** Harmonic Eliashberg functions and superconducting properties calculated in QE with a 16×16×16 *k*-grid, 2×2×2 *q*-grid, and σ-smearing of 0.01 Ry at 165 GPa for (a) *I*4/*mmm*-$YH_4$ and (b) *Imm*2-$YH_7$. "A–D" refers to the Allen–Dynes formula.[30]

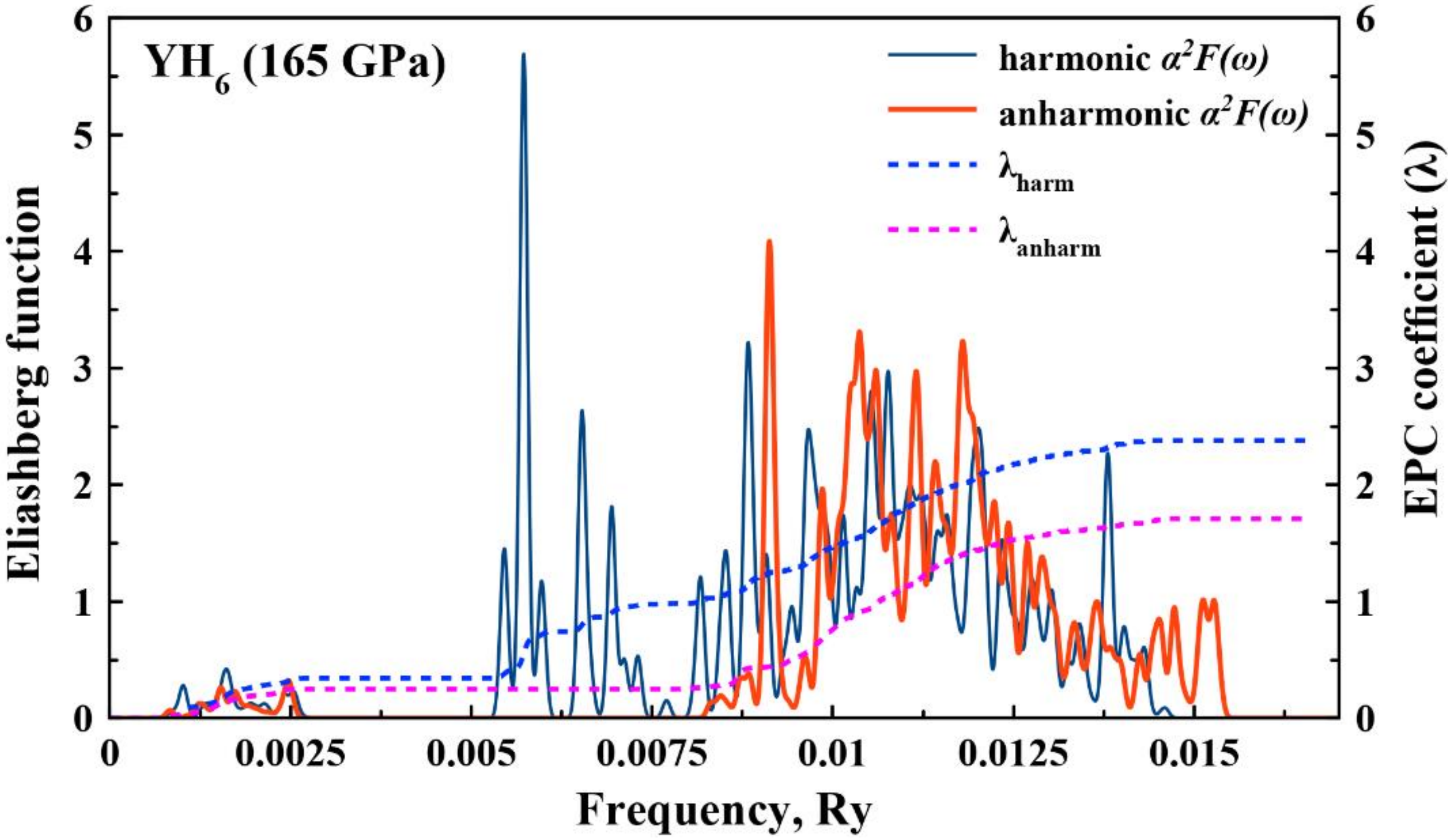

**Figure S16.** Harmonic and anharmonic Eliashberg functions of $Im\overline{3}m$-YH6 at 165 GPa. A 60×60×60 grid of *k*-points and 6×6×6 grid of *q*-points was used in the calculations.

**Table S7.** Parameters of the superconducting state of $Im\bar{3}m$-$YH_6$, *I4/mmm*-$YH_4$, and *Imm*2-$YH_7$ at 165 GPa calculated using the isotropic Migdal–Eliashberg equations (E)[29] and Allen–Dynes (A–D)[30] formula. The Eliashberg functions of $YH_4$ and $YH_7$ were calculated using Quantum ESPRESSO package with 4×4×4 *q*-points and 16×16×16 *k*-points grids. Here, γ is the Sommerfeld constant, α is the Allen–Dynes isotopic coefficient.

| Parameter | $Im\bar{3}m$-$YH_6$[a] | | *I4/mmm*-$YH_4$ | *Imm*2-$YH_7$ |
|---|---|---|---|---|
| | $\mu^* = 0.15$–$0.1$ | $\mu^* = 0.195$ | $\mu^* = 0.15$–$0.1$ | $\mu^* = 0.15$–$0.1$ |
| λ | 1.714 | 1.714 | 1.10 | 0.89 |
| $\omega_{log}$, K | 1333 | 1333 | 1082 | 695 |
| $\omega_2$, K | 1671 | 1671 | 1567 | 1287 |
| α | 0.46–0.48 | 0.43 | 0.44–0.48 | 0.42–0.47 |
| $T_c$ (A–D), K | 168–198 | 149 | 74–94 | 32–43 |
| $T_c$ (E), K | 236–247 | 226 | 90–115 | 36–46 |
| $N(E_F)$, states/eV/f.u. | 0.71 | 0.71 | 0.475 | 0.07 |
| $T_c$ ($YD_x$), K | 168–176 | 168 | 54–67 | 22.4–32 |
| Δ(0), meV | 50–54 | 48 | 16–21 | 5–7.2 |
| $\mu_0 H_C(0)$, T | 63–66 | 60 | 16–21 | 1.9–2.7 |
| $\Delta C/T_c$, mJ/mol·K$^2$ | 26.6–26.8 | 26.3 | 10.1–11.3 | 1.1–1.2 |
| γ, mJ/mol·K$^2$ | 9.075 | 9.1 | 4.7 | 0.62 |
| $R_\Delta = 2\Delta(0)/k_B T_C$ | 4.96–5.03 | 4.9 | 4.1–4.3 | 3.74–3.88 |

[a]60×60×60 *k*-grid, 6×6×6 *q*-grid and anharmonic $\alpha^2F(\omega)$ were used; the harmonic approach gives $T_c$ = 272 K (Table S8).

**Table S8.** Numerical solution of the Migdal–Eliashberg equations for harmonic and anharmonic $\alpha^2F(\omega)$ of $YH_6$ at 165 GPa for different values of the Coloumb pseudopotential.

| Coulomb pseudopotential μ* | 0 | 0.1 | 0.13 | 0.15 | 0.195 |
|---|---|---|---|---|---|
| $T_C$, K (anharmonic) | 269[a] | 247 | 240 | 236 | 226[a] |
| $T_C$, K (harmonic) | 294[a] | 272 | 265 | 261 | 251[a] |

[a]Linear extrapolation.

**Table S9.** Additional electronic and superconducting parameters of $YH_4$ and $YH_7$ at μ* = 0.15–0.1 and $YH_6$ at μ* = 0.195 at the pressure of 165 GPa.

| Parameter | $Im\bar{3}m$-$YH_6$ | *I4/mmm*-$YH_4$ | *Imm*2-$YH_7$ |
|---|---|---|---|
| $N(E_F)$, states/eV/f.u. | 0.71 | 0.475 | 0.07 |
| Average Fermi velocity $V_F$,[a] m/s | 5.3×10$^5$ | 3.4–4.0×10$^5$ | 3.1–3.8×10$^5$ |
| London penetration depth $\lambda_L$, nm | 93 | 180–141 | 208–155 |
| Coherence length, Å | 23 | 45–40 | 130–110 |
| Ginzburg–Landau parameter κ | 40 | 40–35 | 16–14.2 |
| Lower critical magnetic field $\mu_0 H_{C1}$, mT | 50 | 13–21 | 7.5–13 |
| Upper critical field $\mu_0 H_{C2}$, T | 60 | 16–21 | 2.7 |
| Clogston–Chandrasekhar paramagnetic limit, T | 582 | 193–260 | 61–88 |
| Maximum critical current density ($J_c$, A/cm$^2$ | 8.6×10$^8$ | 1.2–2.2×10$^8$ | 3–6.5×10$^7$ |

[a]According to formula (S10)

The critical current density $J_c = e n_e V_L$, evaluated using the Landau criterion for superfluidity,[48] $V_L = \min \frac{\varepsilon(p)}{p} \cong \frac{\Delta_0}{\hbar k_F}$, is around 10$^9$ A/cm$^2$ for $YH_6$, much higher than in $H_3S$.[49]

# Equations for calculating $T_C$ and related parameters

To calculate the isotopic coefficient β, the Allen–Dynes interpolation formulas were used:

$$\beta_{McM} = -\frac{dlnT_C}{dlnM} = \frac{1}{2}\left[1 - \frac{1.04(1+\lambda)(1+0.62\lambda)}{[\lambda - \mu^*(1+0.62\lambda)]^2}\mu^{*2}\right] \quad (S2)$$

$$\beta_{AD} = \beta_{McM} - \frac{2.34\mu^{*2}\lambda^{3/2}}{(2.46+9.25\mu^*)\cdot((2.46+9.25\mu^*)^{3/2}+\lambda^{3/2})} - \frac{130.4\cdot\mu^{*2}\lambda^2(1+6.3\mu^*)\left(1-\frac{\omega_{\log}}{\omega_2}\right)\frac{\omega_{\log}}{\omega_2}}{\left(8.28+104\mu^*+329\mu^{*2}+2.5\cdot\lambda^2\frac{\omega_{\log}}{\omega_2}\right)\cdot\left(8.28+104\mu^*+329\mu^{*2}+2.5\cdot\lambda^2\left(\frac{\omega_{\log}}{\omega_2}\right)^2\right)} \quad (S3)$$

where the last two correction terms are usually small (~0.01).

The Sommerfeld constant was found as

$$\gamma = \frac{2}{3}\pi^2 k_B^2 N(E_F)(1+\lambda) \quad (S4)$$

and was applied to estimate the upper critical magnetic field and the superconductive gap in yttrium hydrides by well-known semiempirical equations of the BCS theory (Ref. [50], Equations 4.1 and 5.11), which can be used for $T_C/\omega_{\log} < 0.25$:

$$\frac{\gamma T_C^2}{B_{C2}^2(0)} = 0.168\left[1 - 12.2\left(\frac{T_C}{\omega_{\log}}\right)^2 \ln\left(\frac{\omega_{\log}}{3T_c}\right)\right] \quad (S5)$$

$$\frac{2\Delta(0)}{k_B T_C} = 3.53\left[1 + 12.5\left(\frac{T_C}{\omega_{\log}}\right)^2 \ln\left(\frac{\omega_{\log}}{2T_c}\right)\right] \quad (S6)$$

The lower critical magnetic field was calculated according to the Ginzburg–Landau theory:[47]

$$\frac{H_{C1}}{H_{C2}} = \frac{\ln k}{2\sqrt{2}k^2}, k = \lambda_L/\xi, \quad (S7)$$

where $\lambda_L$ is the London penetration depth:

$$\lambda_L = 1.0541\times10^{-5}\sqrt{\frac{m_e c^2}{4\pi n_e e^2}}. \quad (S8)$$

Here $c$ is the speed of light, $e$ is the electron charge, $m_e$ is the electron mass, and $n_e$ is an effective concentration of charge carriers, evaluated from the average Fermi velocity ($V_F$) in the Fermi gas model:

$$n_e = \frac{1}{e\pi^2}\left(\frac{m_e V_F}{\hbar}\right)^3. \quad (S9)$$

The coherence length ξ was found as $\xi = \sqrt{\hbar/2e(\mu_0 H_{C2})}$ and was used to estimate the average Fermi velocity:

$$V_F = \frac{\pi\Delta(0)}{\hbar}\xi . \qquad \text{(S10)}$$

The average Fermi velocity was also calculated directly from the band structure of *cI*4-$YH_6$ using

$$\langle V_F \rangle = \sqrt{\frac{\sum_k \delta(E_k - E_F) V_k{}^2}{\sum_k \delta(E_k - E_F)}} = \frac{a}{\pi\hbar}\sqrt{\frac{\sum_k \delta(E_k - E_F)(dE_k/dt)^2}{\sum_k \delta(E_k - E_F)}} . \qquad \text{(S11)}$$

where $dE_i/dk$ ($k = -\pi/a \dots \pi/a$) were replaced by $dE_i/dt$ ($t = -1 \dots 1$). In this case the first Brillouin zone was sampled by the very dense *k*-points mesh with a resolution of 2π×0.005 $Å^{-1}$. The obtained values of $V_F$ of $YH_6$ are $9.1\times10^5$, $9.3\times10^5$ and $9.4\times10^5$ m/s at 150, 165, and 180 GPa, respectively.

Electron band mass was found to be $m^* = 0.82m_e$ for $YH_6$ at 150 GPa ($0.69m_e$ in the vicinity of the Γ-point), which is compatible the electron band mass in $H_3S$ [51].

The calculations within the Migdal–Eliashberg approach were made using the following equations (on the imaginary axis):[29]

$$\Delta_n = \frac{\pi}{\beta}\sum_{m=-Max}^{Max} \frac{\lambda(i\omega_n - i\omega_m) - \mu^*\theta(\omega_c - |\omega_m|)}{\sqrt{\omega_m{}^2 Z_m{}^2 + \Delta_m{}^2}}\Delta_m . \qquad \text{(S12)}$$

$$Z_n = 1 + \frac{\pi}{\beta\omega_n}\sum_{m=-Max}^{Max} \frac{\lambda(i\omega_n - i\omega_m)}{\sqrt{\omega_m{}^2 Z_m{}^2 + \Delta_m{}^2}}\omega_m Z_m . \qquad \text{(S13)}$$

$$\lambda(z) = \int_0^{\omega_c} \frac{2\alpha^2 F(\omega)}{\omega^2 - z^2}\omega d\omega . \qquad \text{(S14)}$$

where $\Delta_n$ is the order parameter function, $Z_n$ is the wave function renormalization factor, θ(*x*) is the Heaviside function, $\omega_n = \pi k_B T(2n - 1)$ is the *n*th Matsubara frequency, $\beta = k_B T$, $\mu^*$ is the Coulomb pseudopotential, λ(z) is the electron–phonon pairing kernel, and $\omega_c$ is the cutoff energy. We used $\omega_c$ = 6 Ry with the appropriate number of the Matsubara frequencies.

# Electronic and phonon properties of yttrium hydrides

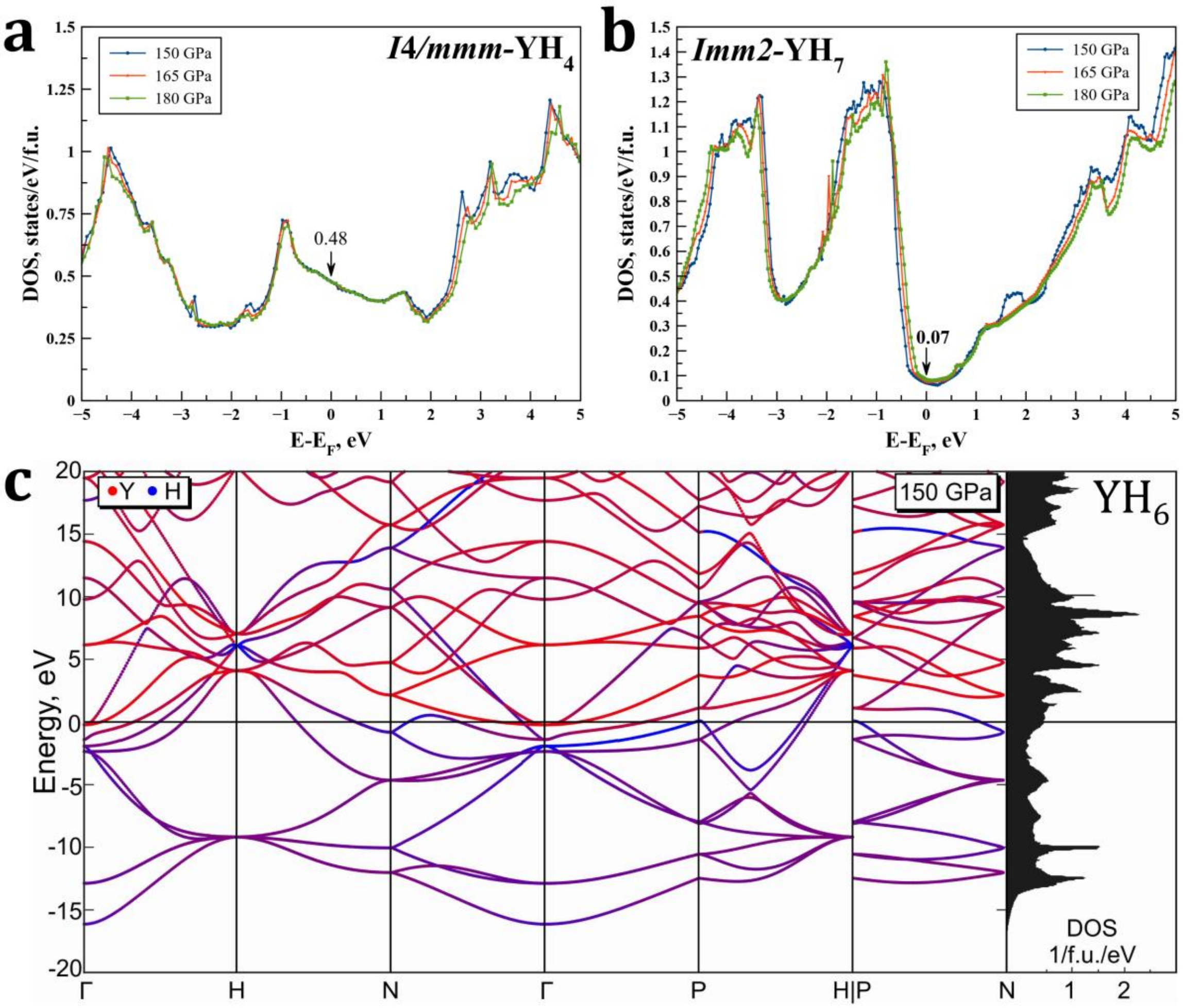

**Figure S17.** Electronic properties of yttrium hydrides: (a) density of electronic states of *I*4/*mmm*-$YH_4$ at $E_F$; (b) density of electronic states of *Imm*2-$YH_7$ at $E_F$; (c) band structure of *cI*4-$YH_6$ at 150 GPa.

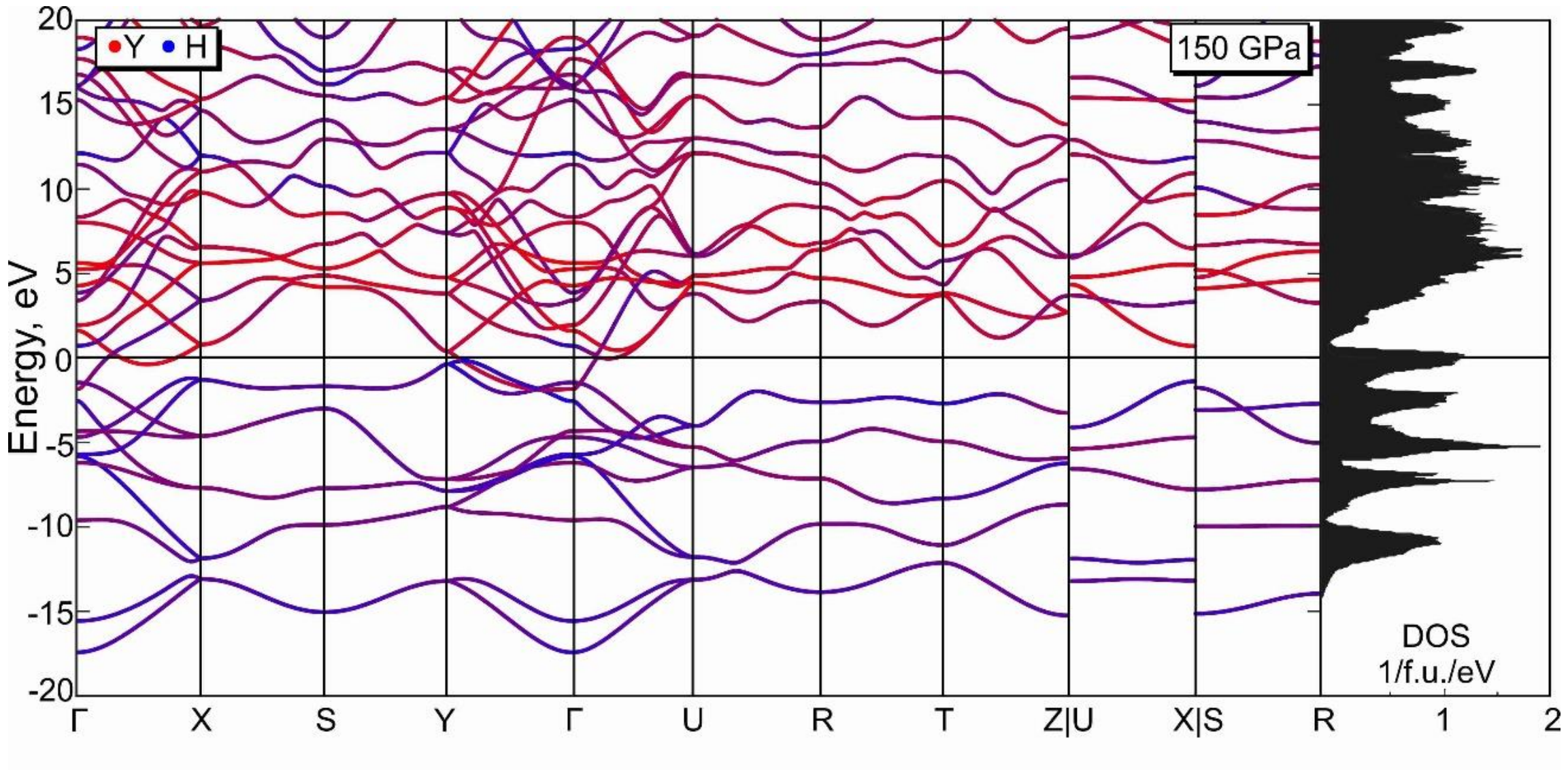

**Figure S18.** Electronic density of states and band structure of *Imm*2-$YH_7$ at 150 GPa.

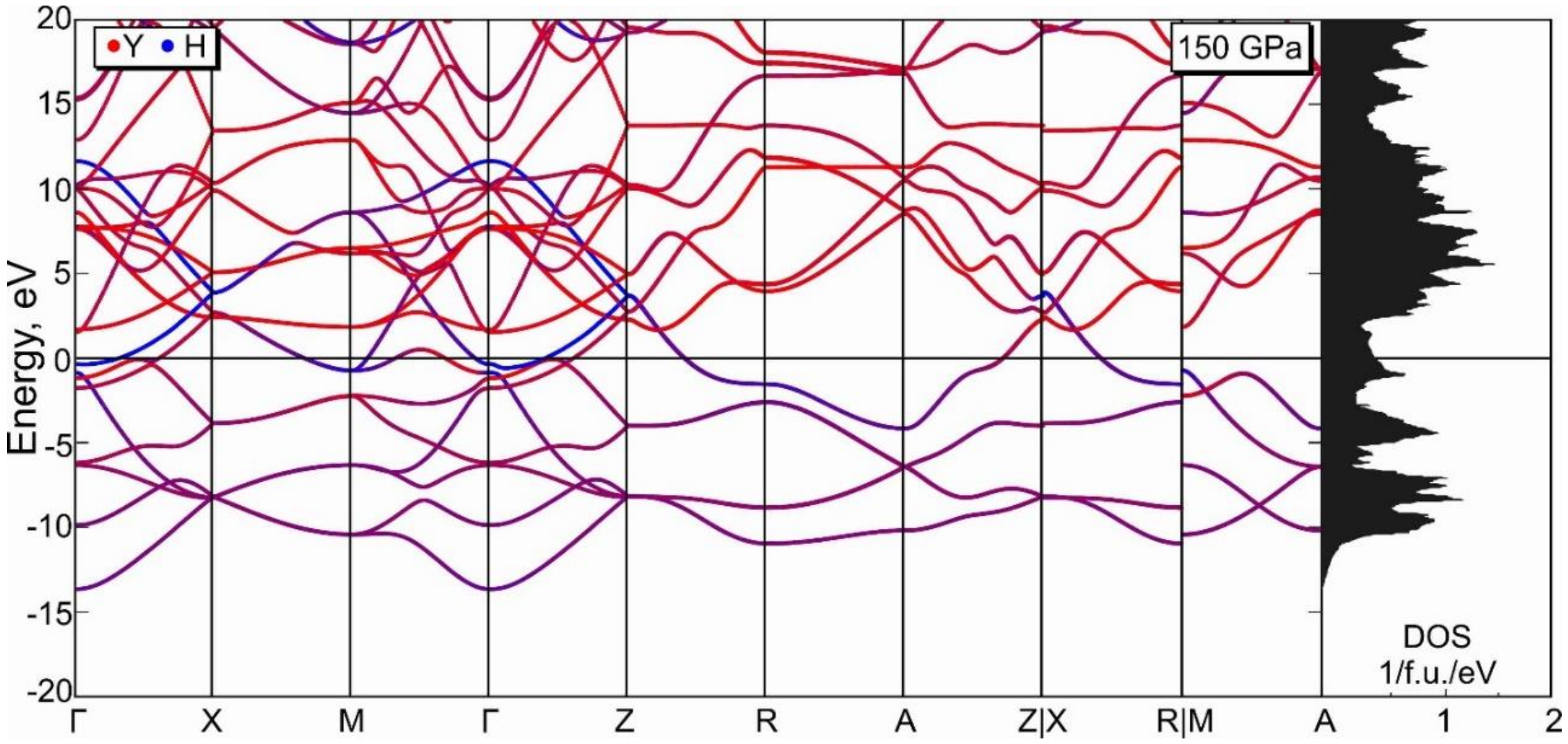

**Figure S19.** Electronic density of states and band structure of *I*4/*mmm*-$YH_4$ at 150 GPa.

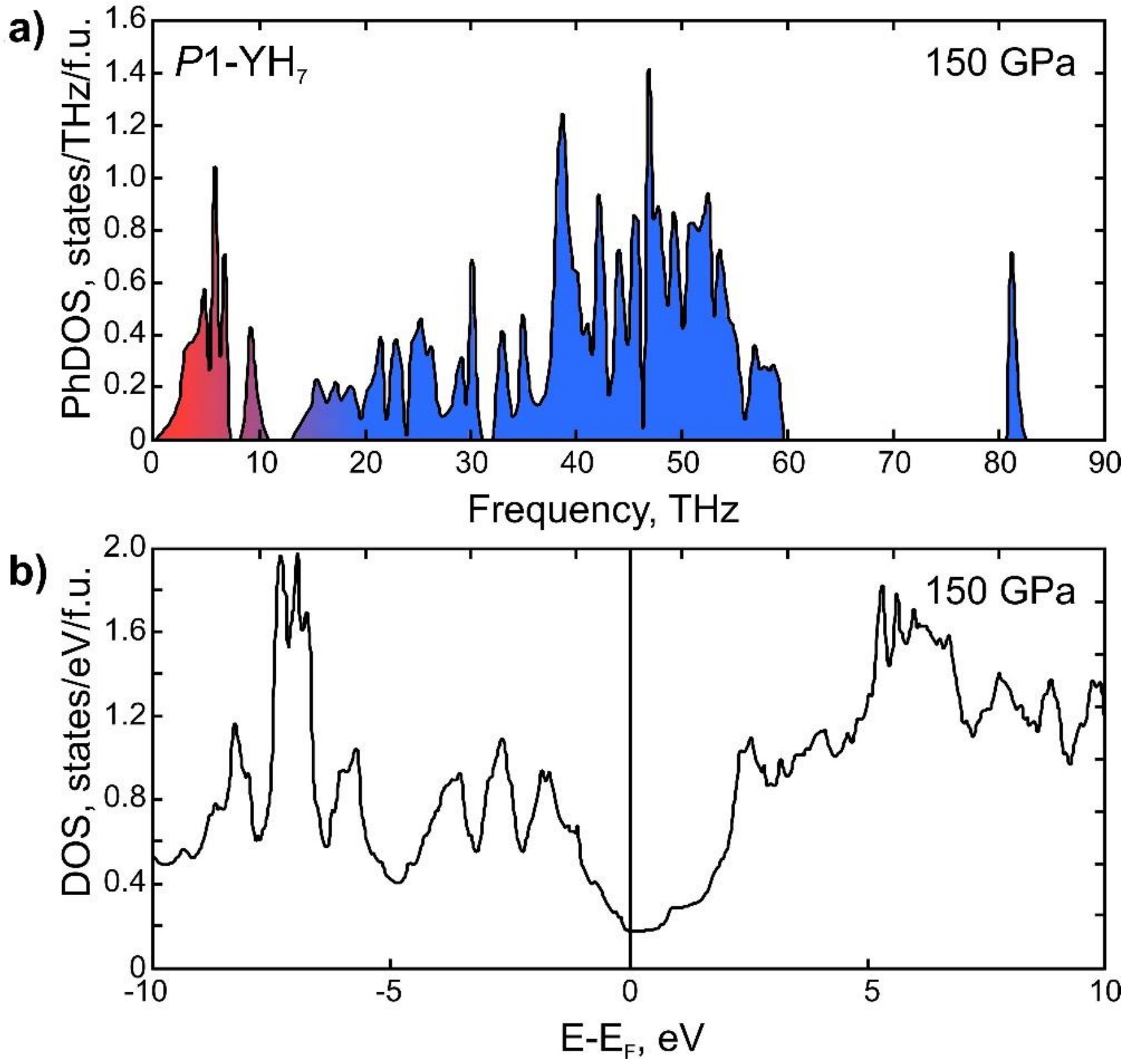

**Figure S20.** (a) Harmonic phonon density of states and (b) electronic density of states of *P*1-$YH_7$ at 150 GPa.

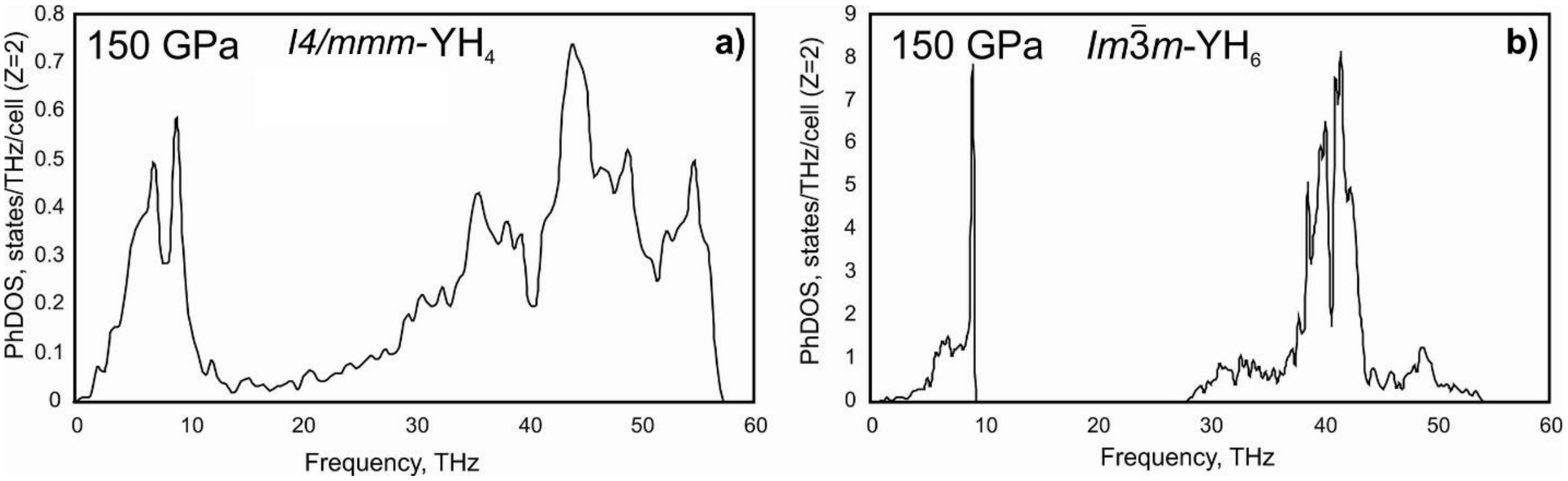

**Figure S21.** Harmonic phonon density of states of (a) $YH_4$ and (b) $YH_6$ at 150 GPa.

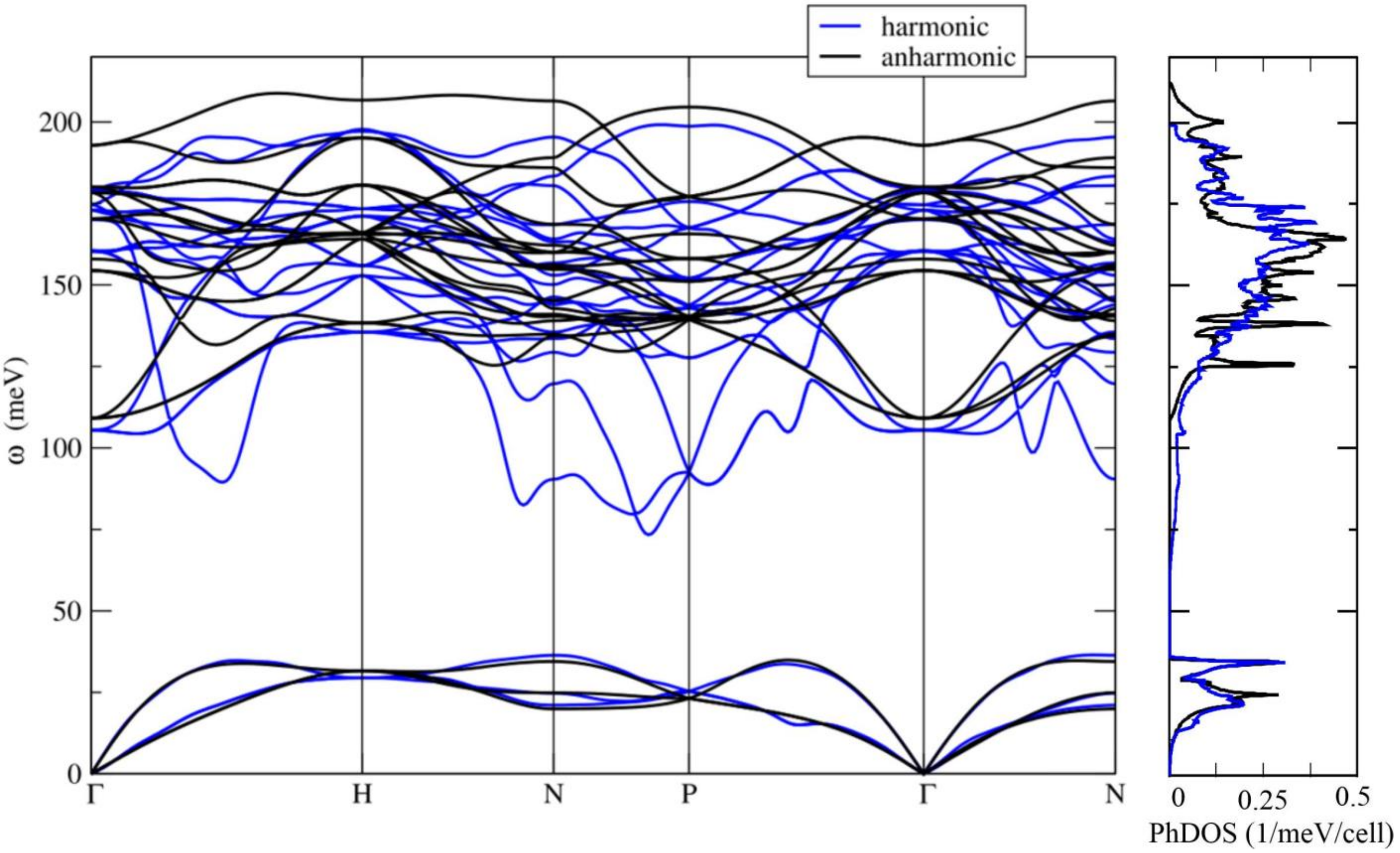

**Figure S22.** Harmonic (blue) and anharmonic (black) phonon band structure and density of states of $Im\bar{3}m$-$YH_6$ at 165 GPa.

# Elastic properties of Y–H phases

The elastic tensors of $YH_4$, $YH_6$ and $YH_7$ were calculated using the stress-strain relations:

$$C_{ij} = \frac{\partial \sigma_i}{\partial \eta_j}, \quad \text{(S15)}$$

where $\sigma_i$ is the $i^{th}$ component of the stress tensor, $\eta_j$ is the $j^{th}$ component of the strain tensor.

The bulk ($B$) and shear ($G$) moduli and Young's modulus ($E$) were calculated in GPa via Voigt-Reuss-Hill averaging. [52,53] Using obtained values of elastic moduli, we calculated the velocities of longitudinal and transverse acoustic waves:

$$v_{LA} = \sqrt{\frac{C_{11}}{\rho}}, \quad v_{TA} = \sqrt{\frac{C_{11} - C_{12}}{2\rho}}, \quad \text{(S16)}$$

where $C_{11}$, $C_{12}$ are elastic constants, ρ is a density of compound. Obtained values allow us to estimate Debye temperature as [54]:

$$\vartheta_D = \frac{h}{k_B}\left[\frac{3n}{4\pi}\left(\frac{N_A \cdot \rho}{M}\right)\right]^{\frac{1}{3}} v_m, \quad \text{(S17)}$$

where $h$, $k_B$, $N_A$ are Planck's, Boltzmann's and Avogadro constants, $v_m$ is an average velocity of acoustic waves calculated by the following formula

$$v_m = \left[\frac{1}{3}\left(\frac{2}{v_{TA}^3} + \frac{1}{v_{LA}^3}\right)\right]^{-1/3}. \quad \text{(S18)}$$

**Table S10.** Elastic and thermodynamic parameters of $Im\bar{3}m$-$YH_6$ ($Z$ = 2).

| Parameter | 150 GPa | 165 GPa | 180 GPa |
|---|---|---|---|
| $a$, Å | 3.602 | 3.573 | 3.546 |
| $V_{DFT}$, Å$^3$ | 46.73 | 45.61 | 44.59 |
| $C_{11}$, GPa | 637 | 870 | 980 |
| $C_{12}$, GPa | 435 | 460 | 482 |
| $C_{44}$, GPa | 109 | 196 | 283 |
| $B$, GPa | 569 | 597 | 648 |
| $G$, GPa | 109 | 196 | 283 |
| $E$, GPa | 307 | 593 | 705 |
| $B/G$ | 5.22 | 3.04 | 2.28 |
| Poisson's ratio η | 0.410 | 0.352 | 0.308 |
| Density ρ, kg/m$^3$ | 6744 | 6910 | 7069 |
| Transverse sound velocity $v_t$, m/s | 4019 | 5326 | 6354 |
| Longitudinal sound velocity $v_l$, m/s | 10291 | 11145 | 12062 |
| Debye temperature $\theta_D$, K | 908 | 1204 | 1438 |

**Table S11.** Elastic and thermodynamic parameters of *I*4/*mmm*-$YH_4$ ($Z$ = 2).

| Parameter | 150 GPa | 165 GPa | 180 GPa |
|---|---|---|---|
| $a$, Å | 2.779 | 2.752 | 2.724 |
| $c$, Å | 5.277 | 5.259 | 5.236 |
| $V_{DFT}$, Å$^3$ | 40.758 | 39.829 | 38.866 |
| $C_{11}$, GPa | 800 | 813 | 856 |
| $C_{12}$, GPa | 425 | 406 | 361 |
| $C_{13}$, GPa | 480 | 508 | 607 |
| $C_{33}$, GPa | 1014 | 1082 | 1035 |
| $C_{44}$, GPa | 229 | 234 | 167 |
| $C_{66}$, GPa | 207 | 233 | 218 |
| $B$, GPa | 592 | 601 | 631 |
| $G$, GPa | 214 | 209 | 183 |
| $E$, GPa | 573 | 562 | 501 |
| $B/G$ | 2.768 | 2.876 | 3.449 |
| Poisson's ratio η | 0.338 | 0.344 | 0.367 |
| Density ρ, kg/m$^3$ | 7570 | 7747 | 7939 |
| Transverse sound velocity $v_t$, m/s | 5330 | 5197 | 4816 |
| Longitudinal sound velocity $v_l$, m/s | 10774 | 10658 | 10508 |
| Debye temperature $\theta_D$, K | 1115 | 1097 | 1028 |

**Table S12.** Elastic and thermodynamic parameters of *Imm*2-$YH_7$ ($Z$ = 2).

| Parameter | 150 GPa | 165 GPa | 180 GPa |
|---|---|---|---|
| $a$, Å | 3.281 | 3.258 | 3.235 |
| $b$, Å | 3.402 | 3.369 | 3.340 |
| $c$, Å | 4.676 | 4.632 | 4.590 |
| $V_{DFT}$, Å$^3$ | 52.19 | 50.85 | 49.59 |
| $C_{11}$, GPa | 821 | 847 | 910 |
| $C_{12}$, GPa | 369 | 374 | 436 |
| $C_{13}$, GPa | 480 | 505 | 559 |
| $C_{22}$, GPa | 896 | 923 | 958 |
| $C_{23}$, GPa | 310 | 334 | 351 |
| $C_{33}$, GPa | 840 | 881 | 926 |
| $C_{44}$, GPa | 138 | 140 | 142 |
| $C_{55}$, GPa | 122 | 136 | 143 |
| $C_{66}$, GPa | 328 | 407 | 425 |
| $B$, GPa | 541 | 563 | 608 |
| $G$, GPa | 195 | 212 | 218 |
| $E$, GPa | 523 | 566 | 583 |
| $B/G$ | 2.766 | 2.649 | 2.793 |
| Poisson ratio η | 0.338 | 0.332 | 0.340 |
| Density ρ, kg/m$^3$ | 5920 | 6075 | 6228 |
| Transverse sound velocity $v_t$, m/s | 5760 | 5920 | 5113 |
| Longitudinal sound velocity $v_l$, m/s | 11650 | 11810 | 12010 |
| Debye temperature $\theta_D$, K | 1300 | 1345 | 1356 |

# References

1. Prescher, C. & Prakapenka, V. B. DIOPTAS: a program for reduction of two-dimensional X-ray diffraction data and data exploration. *High Pressure Research* **35**, 223–230 (2015).
2. Petříček, V., Dušek, M. & Palatinus, L. Crystallographic Computing System JANA2006: General features. *Zeitschrift für Kristallographie - Crystalline Materials* **229**, 345–352 (2014).
3. Bail, A. L. Whole powder pattern decomposition methods and applications: A retrospection. *Powder Diffraction* **20**, 316–326 (2005).
4. Prakapenka, V. B. *et al.* Advanced flat top laser heating system for high pressure research at GSECARS: application to the melting behavior of germanium. *High Pressure Research* **28**, 225–235 (2008).
5. Akahama, Y. & Kawamura, H. Pressure calibration of diamond anvil Raman gauge to 410 GPa. *J. Phys.: Conf. Ser.* **215**, 012195 (2010).
6. Holtgrewe, N., Greenberg, E., Prescher, C., Prakapenka, V. B. & Goncharov, A. F. Advanced integrated optical spectroscopy system for diamond anvil cell studies at GSECARS. *High Pressure Research* **39**, 457–470 (2019).
7. Eremets, M. I. Megabar high-pressure cells for Raman measurements. *Journal of Raman Spectroscopy* **34**, 515–518 (2003).
8. Parvanov, V. M. *et al.* Materials for hydrogen storage: structure and dynamics of borane ammonia complex. *Dalton Trans.* 4514–4522 (2008) doi:10.1039/B718138H.
9. Dew-Hughes, D. Flux pinning mechanisms in type II superconductors. *The Philosophical Magazine: A Journal of Theoretical Experimental and Applied Physics* **30**, 293–305 (1974).
10. Oganov, A. R. & Glass, C. W. Crystal structure prediction using ab initio evolutionary techniques: Principles and applications. *J. Chem. Phys.* **124**, 244704 (2006).
11. Oganov, A. R., Lyakhov, A. O. & Valle, M. How Evolutionary Crystal Structure Prediction Works—and Why. *Acc. Chem. Res.* **44**, 227–237 (2011).
12. Lyakhov, A. O., Oganov, A. R., Stokes, H. T. & Zhu, Q. New developments in evolutionary structure prediction algorithm USPEX. *Computer Physics Communications* **184**, 1172–1182 (2013).
13. Bushlanov, P. V., Blatov, V. A. & Oganov, A. R. Topology-based crystal structure generator. *Computer Physics Communications* **236**, 1–7 (2019).
14. Hohenberg, P. & Kohn, W. Inhomogeneous electron gas. *Phys Rev* **136**, B864–B871 (1964).
15. Kohn, W. & Sham, L. J. Self-consistent equations including exchange and correlation effects. *Phys Rev* **140**, A1133–A1138 (1965).
16. Perdew, J. P., Burke, K. & Ernzerhof, M. Generalized gradient approximation made simple. *Physical review letters* **77**, 3865–3868 (1996).
17. Blöchl, P. E. Projector augmented-wave method. *Phys. Rev. B* **50**, 17953–17979 (1994).
18. Kresse, G. & Joubert, D. From ultrasoft pseudopotentials to the projector augmented-wave method. *Phys. Rev. B* **59**, 1758–1775 (1999).
19. Kresse, G. & Furthmüller, J. Efficient iterative schemes for ab initio total-energy calculations using a plane-wave basis set. *Phys. Rev. B* **54**, 11169–11186 (1996).
20. Kresse, G. & Hafner, J. Ab initio molecular dynamics for liquid metals. *Phys. Rev. B* **47**, 558–561 (1993).
21. Kresse, G. & Hafner, J. Ab initio molecular-dynamics simulation of the liquid-metal amorphous-semiconductor transition in germanium. *Phys. Rev. B* **49**, 14251–14269 (1994).
22. Semenok, D. V., Kvashnin, A. G., Kruglov, I. A. & Oganov, A. R. Actinium Hydrides AcH10, AcH12, and $AcH_{16}$ as High-Temperature Conventional Superconductors. *J. Phys. Chem. Lett.* **9**, 1920–1926 (2018).

23. Kvashnin, A. G., Semenok, D. V., Kruglov, I. A., Wrona, I. A. & Oganov, A. R. High-Temperature Superconductivity in Th-H System at Pressure Conditions. *ACS Applied Materials & Interfaces* **10**, 43809–43816 (2018).
24. Togo, A. & Tanaka, I. First principles phonon calculations in materials science. *Scripta Materialia* **108**, 1–5 (2015).
25. Togo, A., Oba, F. & Tanaka, I. First-principles calculations of the ferroelastic transition between rutile-type and CaCl2-type SiO2 at high pressures. *Phys. Rev. B* **78**, 134106 (2008).
26. Giannozzi, P. *et al.* QUANTUM ESPRESSO: a Modular and Open-Source Software Project for Quantum Simulations of Materials. *Journal of Physics: Condensed Matter* **21**, 395502 (2009).
27. Giannozzi, P. *et al.* Advanced capabilities for materials modelling with Quantum ESPRESSO. *J. Phys.: Condens. Matter* **29**, 465901 (2017).
28. Baroni, S., de Gironcoli, S., Dal Corso, A. & Giannozzi, P. Phonons and Related Crystal Properties from Density-Functional Perturbation Theory. *Reviews of modern Physics* **73**, 515–562 (2001).
29. Eliashberg, G. M. Interactions between Electrons and Lattice Vibrations in a Superconductor. *JETP* **11**, 696–702 (1959).
30. Allen, P. B. & Dynes, R. C. Transition temperature of strong-coupled superconductors reanalyzed. *Phys. Rev. B* **12**, 905–922 (1975).
31. Lüders, M. *et al.* Ab initio theory of superconductivity. I. Density functional formalism and approximate functionals. *Phys. Rev. B* **72**, 024545 (2005).
32. Marques, M. A. L. *et al.* Ab initio theory of superconductivity. II. Application to elemental metals. *Phys. Rev. B* **72**, 024546 (2005).
33. Kruglov, I. A. *et al.* Superconductivity of LaH10 and LaH16 polyhydrides. *Phys. Rev. B.* **101**, 024508 (2020).
34. Akashi, R. & Arita, R. Density functional theory for superconductors with particle-hole asymmetric electronic structure. *Phys. Rev. B* **88**, 014514 (2013).
35. Akashi, R., Kawamura, M., Tsuneyuki, S., Nomura, Y. & Arita, R. First-principles study of the pressure and crystal-structure dependences of the superconducting transition temperature in compressed sulfur hydrides. *Phys. Rev. B* **91**, 224513 (2015).
36. Migdal, A. B. Interaction between electrons and lattice vibrations in a normal metal. *JETP* **34**, 996–1001 (1958).
37. Scalapino, D. J. The Electron Phonon Interaction and Strong Coupling. in *Superconductivity: Part 1 (In Two Parts)* (ed. Parks, R. D.) vol. 1 (Dekker, New York, 1969).
38. Schrieffer, J. R. *Theory Of Superconductivity*. vol. 1 (Perseus Books, 1999).
39. Pines, D. *Elementary Excitations in Solids : Lectures on Phonons, Electrons, and Plasmons*. (Perseus, 1999).
40. Rath, J. & Freeman, A. J. Generalized magnetic susceptibilities in metals: Application of the analytic tetrahedron linear energy method to Sc. *Phys. Rev. B* **11**, 2109–2117 (1975).
41. Errea, I., Calandra, M. & Mauri, F. Anharmonic free energies and phonon dispersions from the stochastic self-consistent harmonic approximation: Application to platinum and palladium hydrides. *Phys. Rev. B* **89**, 064302 (2014).
42. Grishakov, K. S., Degtyarenko, N. N. & Mazur, E. A. Electron, Phonon, and Superconducting Properties of Yttrium and Sulfur Hydrides under High Pressures. *J. Exp. Theor. Phys.* **128**, 105–114 (2019).
43. Heil, C., di Cataldo, S., Bachelet, G. B. & Boeri, L. Superconductivity in sodalite-like yttrium hydride clathrates. *Phys. Rev. B* **99**, 220502 (2019).
44. Li, Y. *et al.* Pressure-stabilized superconductive yttrium hydrides. *Scientific Reports* **5**, 09948 (2015).
45. Liu, H., Naumov, I. I., Hoffmann, R., Ashcroft, N. W. & Hemley, R. J. Potential high-Tc superconducting lanthanum and yttrium hydrides at high pressure. *PNAS* **114**, 6990–6995 (2017).

46. Bergara, A. Anharmonic effects and optical spectra in superconducting hydrogen. in *Proceedings of 56th EHPRG Meeting* (2018).
47. Ginzburg, V. L. & Landau, L. D. On the Theory of superconductivity. *Zh.Eksp.Teor.Fiz.* **20**, 1064–1082 (1950).
48. Landau, L. Theory of the Superfluidity of Helium II. *Phys. Rev.* **60**, 356–358 (1941).
49. Drozdov, A. P., Eremets, M. I., Troyan, I. A., Ksenofontov, V. & Shylin, S. I. Conventional superconductivity at 203 kelvin at high pressures in the sulfur hydride system. *Nature* **525**, 73–76 (2015).
50. Carbotte, J. P. Properties of boson-exchange superconductors. *Rev. Mod. Phys.* **62**, 1027–1157 (1990).
51. Quan, Y. & Pickett, W. E. Van Hove singularities and spectral smearing in high-temperature superconducting H3S. *Phys. Rev. B* **93**, 104526 (2016).
52. Hill, R. The Elastic Behaviour of a Crystalline Aggregate. *Proc. Phys. Soc. A* **65**, 349 (1952).
53. Anderson, O. L. A simplified method for calculating the debye temperature from elastic constants. *Journal of Physics and Chemistry of Solids* **24**, 909–917 (1963).
54. Ravindran, P. *et al.* Density functional theory for calculation of elastic properties of orthorhombic crystals: Application to TiSi2. *Journal of Applied Physics* **84**, 4891–4904 (1998).